\documentclass{aa}  

\usepackage{natbib}
\usepackage{graphicx}
\usepackage{babel}
\usepackage{caption}
\usepackage{subfig}
\usepackage[varg]{txfonts}
\usepackage{rotating}
\usepackage{dcolumn}
\usepackage{fmtcount}
\usepackage{longtable, booktabs}
\usepackage{array,multirow}
\usepackage{color}
\usepackage{aas_macros}
\usepackage{multirow}

\usepackage{dcolumn} 

\bibpunct{(}{)}{;}{a}{}{,}

\usepackage{hyperref}

\usepackage{xcolor}
\usepackage{ulem}
\definecolor{dark-red}{rgb}{0.5,0.15,0.15}
\definecolor{dark-blue}{rgb}{0.15,0.15,0.5}
\definecolor{medium-blue}{rgb}{0,0,0.5}
\definecolor{medium-red}{rgb}{1,0,0}
\hypersetup{
    colorlinks={true}, linkcolor={medium-red},
    citecolor={dark-blue}, urlcolor={medium-blue}
}

\newcommand\mc[1]{\multicolumn{1}{c}{#1}} 

\newcommand{\teff}{\ensuremath{T_{\mathrm{eff}}}\xspace}
\newcommand{\kms}{\ensuremath{\rm{km}\,s^{-1}}\xspace}
\newcommand{\logg}{\ensuremath{\log g}\xspace}
\newcommand{\feh}{{\rm{[Fe/H]}}\xspace}
\newcommand{\fehe}{\ensuremath{\rm{[Fe/H]_e}}\xspace}
\newcommand{\afe}{{\rm{[\ensuremath{\alpha}/Fe]}}\xspace}

\newcommand{\mfeh}{\ensuremath{\langle\rm{[Fe/H]}\rangle}\xspace}

\defcitealias{jain20}{J20}               
\newcommand{\RJ}{\citetalias{jain20}\xspace}

\defcitealias{husser16}{H16}             
\newcommand{\TOH}{\citetalias{husser16}\xspace}

\defcitealias{allende-prieto18}{AP18}            
\newcommand{\AP}{\citetalias{allende-prieto18}\xspace}

\usepackage{xcolor,ulem}
\newcommand\hl{\bgroup\markoverwith{\textcolor{yellow}{\rule[-.5ex]{2pt}{2.5ex}}}\ULon} 
\newcommand\hlM{\bgroup\markoverwith{\textcolor{pink}{\rule[-.5ex]{2pt}{2.5ex}}}\ULon} 
\newcommand\hlD{\bgroup\markoverwith{\textcolor{green}{\rule[-.5ex]{2pt}{2.5ex}}}\ULon} 
\newcommand{\notep}[1]{{#1}}  

\usepackage{amstext}

\begin{document} 

   \title{Prospects of measuring a metallicity trend and spread in globular clusters from low-resolution spectroscopy \thanks{The Table with the results for each individual star is only available in electronic format the CDS via anonymous ftp to cdsarc.u-strasbg.fr (130.79.128.5) or via \url{http://cdsweb.u-strasbg.fr/cgi-bin/qcat?J/A+A/}}}
\titlerunning{Metallicity trend and spread in globular clusters from low-resolution spectroscopy}
   
   \author{ M.Baratella\inst{\ref{aip}}
   \and Deepthi S. Prabhu\inst{\ref{iia}, \ref{pondi}}
   \and L. Lima\inst{\ref{brasil}}
   \and P. Prugniel\inst{\ref{cral}}
   }
   \institute{Leibniz-Institut f\"ur Astrophysik Potsdam (AIP), An der Sternwarte 16, 14482 Potsdam, Germany \label{aip}\\
\email{mbaratella@aip.de}
    \and
    Indian Institute of Astrophysics, Koramangala II Block, Bangalore-560034, India \label{iia}
    \and 
    Pondicherry University, R.V. Nagar, Kalapet, 605014, Puducherry, India \label{pondi}
    \and
    NAT - Universidade Cidade de São Paulo/Universidade Cruzeiro do Sul, Rua Galvão Bueno 868, São Paulo-SP, 01506-000, Brazil\label{brasil}
    \and
    Centre de Recherche Astrophysique de Lyon (CRAL, CNRS, UMR 5574), Université Lyon 1, Ecole Nationale Supérieure de Lyon, Université de Lyon, France  \label{cral}
   }
   \date{Received xxxx; accepted xxxx}


  \abstract
      {
        The metallicity spread, or the metallicity trend along the evolutionary sequence of a globular cluster, is a rich source of information to help understand the cluster physics (e.g. multiple populations) and stellar physics (e.g. atomic diffusion). Low-resolution integral-field-unit spectroscopy in the optical with the MUSE spectrograph is an attractive prospect if it can provide these diagnostics because it allows us to efficiently extract spectra of a large fraction of the cluster stars with only a few telescope pointings. 
            }
      {We investigate the possibilities of full-spectrum fitting to derive stellar parameters and chemical abundances at low spectral resolution (R$\sim$2\,000). 
   }
   {We reanalysed 1584 MUSE spectra of 1061 stars above the turn-off of NGC\,6397 using FERRE and employing two different synthetic libraries. 
   }
   {We derive the equivalent iron abundance \fehe for fixed values of \afe (solar or enhanced). We find that (i) the interpolation schema and grid mesh are not critical for the precision, metallicity spread, and trend; 
\notep{(ii) with the two considered grids, \fehe increases by $\sim 0.2$\,dex along the sub-giant branch, starting from the turn-off of the main sequence;}
     (iii) restricting the wavelength range to the optical  
decreases the precision significantly; and (iv) the precision obtained with the synthetic libraries is lower than the precision obtained previously with empirical libraries.
}
   {
     Full-spectrum fitting provides reproducible results that are robust to the choice of the reference grid of synthetic spectra and to the details of the analysis.
     The \fehe increase along the sub-giant branch is in stark contrast with the nearly constant iron abundance previously found with empirical libraries.
     The precision of the measurements (0.05 dex on \fehe) is currently not sufficient to assess the intrinsic chemical abundance spreads, but this may change with deeper observations.
     Improvements of the synthetic spectra are still needed to deliver the full possibilities of full-spectrum fitting.
   }


   \maketitle

\section{Introduction}\label{sec:introduction}

Globular clusters (GCs) are studied for their own sakes, for instance, to investigate the characteristics of their stellar population and infer their formation and evolution processes. They are also benchmarks for probing theories of stellar physics, such as the variations in the surface chemical composition of stars with respect to their initial composition as expected from atomic diffusion \citep{chapman1917}.

Significant abundance spreads were documented in \citet{gratton12} or \citet{meszaros15}, for  example, with amplitudes larger by more than 0.1 dex for the $\alpha$-elements. With a few exceptions such as $\omega$ Cen and Terzan 5, the iron (\feh) spreads are in contrast lower than the observational uncertainties in general (about 0.05\,dex; \citealt{carretta2009}). Some abundance trends (i.e. systematic change along the temperature sequence of the cluster) were also suggested \citep[in NGC\,6397]{korn2007,lind2008,nordlander2012}. These studies are at the precision limit of the measurements, and a number of observational or modelling effects can mislead the interpretation. 

In \notep{these} investigations, the authors analysed observations with high signal-to-noise ratio (S/N) and high spectral resolution of as many stars as possible. However, these observations are expensive because the multiplexing capability of high-resolution spectrographs is limited and exposure times are long, and they are hampered by crowding in the central regions of the clusters\notep{. The final samples are therefore often restricted to typically a dozen stars in the outskirts of the cluster}. Low-resolution integral-field-unit (IFU) spectroscopy may be considered as an alternative. It benefits from a large multiplexing and from the development of crowded-field spectroscopy \citep{kamann13}. 

The MUSE IFU spectrograph, attached to the ESO Very Large Telescope \citep{2010bacon}, allows observing a significant fraction of a Galactic GC in a single telescope pointing. It has been used by \citet[hereafter H16]{husser16} to observe NGC\,6397, \notep{where the parameters of about $1600$ stars above the turn-off (TO) point of the main sequence were measured.
Because of the low spectral resolution (R$\sim$2000), individual spectral lines from different chemical species are blended, and classical methods are not applicable. Therefore, the authors used full-spectrum fitting \citep[see e.g.][]{koleva09}, where all the spectral bins are compared at once to reference spectra to derive the atmospheric parameters such as effective temperature (\teff), surface gravity (\logg), and \feh. This has demonstrated that low-resolution IFU spectroscopy can be effective at analysing large samples of stars in a GC with good internal precision. 
\TOH found a \feh trend of $\sim 0.2$\,dex along the effective temperature sequence, but when the same dataset was reanalysed by \citet[hereafter J20]{jain20}, the iron abundance was found to be nearly constant ($\feh \sim -2.0$\,dex). Both studies used the full-spectrum fitting technique to derive the parameters, and the origins of the different results lie in the details of the analyses. Specifically, while \TOH was using synthetic spectra from the G\"ottingen Spectral Library\footnote{\url{https://phoenix.astro.physik.uni-goettingen.de}} (hereafter GSL; \citealt{2013Husser}) as reference, \RJ used empirical libraries (ELODIE; \citealt{2001prugniel,wu2011}, and MILES; \citealt{sanchez-blazquez2006}).}
\notep{We also stress that with full-spectrum fitting}, the derived iron abundance depends on the abundance pattern inscribed in the set of reference spectra. For this reason, we call it an \textit{\textup{equivalent}} iron abundance, hereafter noted \fehe, whose relation to the true abundance has to be considered with care.

To assess the external accuracy of these results, we need to understand the systematics caused by the analysis methods, such as those due to the interpolation performed within the spectral library and those due to the adopted libraries.
Comparing the results obtained with different libraries and methods may give a lower limit of this second source of errors.

As a step in this direction, we reanalyse here the set of MUSE observations previously used by \TOH and \RJ. We use a different analysis tool and  different sets of reference models to explore the prospects of using full-spectrum fitting methods to derive accurate parameters from the analysis of low-resolution stellar spectra.
The paper is organised as follows: in Sect.~\ref{sec:data} we present the observational material, and in Sect.~\ref{sec:methods} we describe the analysis method based on FERRE\footnote{\url{https://github.com/callendeprieto/ferre}} \citep{allende-prieto15}. Sect.~\ref{sec:anal_gsl} presents the analysis with the GSL grid (the same grid as was used by \TOH), hence allowing us to compare the results and investigate the effect of the interpolation scheme and of other details of the setup. In Sect.~\ref{sec:anal_ap} we carry out the analysis with the \citet[hereafter AP18]{allende-prieto18} grid of models to investigate the effect of the choice of model and of the mesh size. We finally discuss the different aspects that affect the ability of measuring metallicity trends and spreads in Sect.\,\ref{sec:discussion}, and we draw conclusions in Sect.\,\ref{sec:conclusion}.

\section{Observational material}\label{sec:data}

NGC\,6397 was observed during the MUSE commissioning between July 26 and August 3, 2014 (ESO ID program \texttt{60.A-9100(C)}). The observations consist of a 5x5 square mosaic (two tiles are missing) and reach about 3.5 arcmin from the cluster centre on its diagonal. They are based on 127 pointings with exposure times of about 60\,s to avoid saturation of the brightest stars for a total integration time of 95\,min. Each pointing covers a field of 1x1 arcmin, and the seeing is between 0.6 and 1 arcsec. All the data have already been reduced by the MUSE consortium using the official pipeline. The MUSE spectra cover a wide wavelength range, $\lambda\lambda =  470 - 950$\,nm, with a resolution R $= \lambda/\Delta\lambda \sim2\,000$, where $\lambda$ is the wavelength, and $\Delta\lambda$ the full width at half maximum (FWHM) of an unresolved line.

The spectra were extracted by \TOH (see that paper for further details) and were downloaded from the website set for commissioning data\footnote{\url{http://muse-vlt.eu/science/globular-cluster-ngc-6397/}}. 
The typical S/N of stars at the TO is $\sim 50$ on average, and the brightest giants reach S/N\,$\sim\,200$. The table with photometry and measurements made in \TOH were kindly provided by Tim-Oliver Husser (priv. comm.). 
We corrected the spectra for the telluric absorption by dividing each stellar spectrum by the corresponding telluric spectrum determined by \TOH. We then reduced the spectra to the rest-frame velocity by changing the world coordinate system (WCS), using the mean velocity of the cluster. Because the sampling is not exactly the same in all the spectra (starting wavelength at $4749.75$ or $4750$\,\AA), we rebinned all of them to the same WCS to simplify the analysis with FERRE (see Sect.\,\ref{sec:methods}). 
The normalisation of the spectra to their continuum was performed during the analysis in order to employ the same algorithm and parameters for the observations and the models.
\notep{The weight of the individual spectral bins was computed assuming a constant S/N throughout the spectrum. We experimented with various weighting schemes and retained the scheme that provided the best precision.}

These commissioning observations are based on short observing time and moderate image quality (they were obtained before the ground-layer adaptive optics entered routine use to improve the image quality). Moreover, since that early epoch, the data reduction software has been improved. Still, this public collection of reduced spectra is unique in its the possibility of comparing different approaches for measuring stellar atmospheric parameters.

\subsection{Selected sample and surface gravities}
\label{sec:data_logg}

\TOH used this observational material to determine the atmospheric parameters of 4132 stars from 5881 spectra (because the individual tiles of the mosaic overlap, some stars were observed more than once) with S/N $ > 20$. With this S/N cut-off, the formal errors on \teff and \fehe are lower than 100\,K and 0.16\,dex for stars along the giant and sub-giant (SG) branches, and near the TO, as discussed by \TOH. Later, \RJ restricted this sample further to the cluster members (based on the radial velocity), excluded the hot stars (restricting to $\teff < 7000$\, K), and finally selected only the stars with $\logg < 4.2$. This resulted in a sample of 1587 spectra for 1063 stars.

We adopt the values of \logg computed by \TOH using the photometry from \citet{anderson2008} and isochrones from \citet{bressan12}. 
\RJ has shown that fitting \logg at the same time as \teff and \fehe produces results consistent with the photometric \logg and does not affect the observed \notep{iron abundance} trend. Adopting this photometric \logg simplifies the comparison with the earlier studies. 

In the \RJ sample, three of the 1587 stars have \logg$<1$, but the two model grids that we use (GSL and \AP) start at \logg~$=1$ because the slice at \logg$=0$ is incomplete (see Sect. \ref{sec:anal_gsl} for further details). Thus, we excluded them from our analysis, and the final sample consists of 1584 spectra for 1061 stars.\\

\subsection{Injection of the line-spread function}

The full-spectrum fitting method we used involves comparing each spectral bin of an observed spectrum to the corresponding bin of model spectra and searching for the physical parameters that provide the best match (in a $\chi^2$ sense). Because models are computed at a high resolution, they must be transformed to have the same line-spread function (LSF) as the observations. This process is called LSF injection, and the function used in the transformation is called relative LSF. 

The mean LSF of the MUSE spectra varies with wavelength, and it
varies by 0.1\,\AA\, across the 24 spectrographs that form MUSE. \TOH has pointed out the difficulty of accurately matching the LSF for each individual spectrum,
and following them and \RJ, we neglected the variation between the different spectrograph.
This simplification is justified a posteriori in Sect.~\ref{sec:lsf_effect} by testing the effect of the LSF width and centring on the uncertainties.

To model its variation across the wavelength range, we defined the LSF in two segments, and we used a piece-wise convolution similar to the implementation in ULySS\footnote{\url{http://ulyss.univ-lyon1.fr/}} \citep{koleva09}. As the GSL and \AP models have the same spectral resolution \notep{(R $= 10\,000$)}, we adopted the same relative Gaussian LSF for both: at $\lambda = 4750\,\AA$, the FWHM is $2.75\,\AA$, at $\lambda =  7000\,\AA$, it is $2.45\,\AA$, and at $\lambda =  9300\,\AA$, it is $2.50\,\AA$. Moreover, we noticed a small shift in the NIR range, for which the spectra were corrected by adding a 0.20\,\AA\, shift at $\lambda = 9300\,\AA$.
Each model spectrum was convolved with these three LSFs, and the final model was obtained by interpolating linearly in wavelength between the first and second spectra for wavelengths smaller than $7000\,\AA$, and between the second and third spectra for larger wavelengths.

\begin{table}[!htb]
\caption{Parameter space of the grids. }
\centering
\setlength{\tabcolsep}{1pt}
\begin{tabular}{@{\hspace{0.5em}}l@{\hspace{4em}}c@{\hspace{4em}}rl@{\hspace{0.5em}}}
\toprule
Variable & Range & Step \\
\midrule
\multicolumn{4}{c}{\textbf{GSL grids}, \afe = 0.4 \& 0\,dex}\\
\hline 
\teff   &   5100 : 7000    &   100 &K\\
\logg   &   1 : 6   &   0.5 &dex\\
\feh    &   $-3$ : 0  &   0.5 &dex\\
\midrule
\multicolumn{4}{c}{\textbf{\AP fine grids}, \afe = 0.5 \& 0\,dex, $\xi = 1$ \& $2$\,\kms}\\
\midrule
\teff   &    4000 : 7000 &    250 &K\\
\logg   &   1 : 5   &    0.5 &dex\\
\feh    &   $-3$ : 0    &   0.25 &dex\\
\midrule
\multicolumn{4}{c}{\textbf{\AP coarse grid}, \afe = 0.5\,dex, $\xi = 1$\,\kms}\\
\midrule
\teff   &    4000 : 7000 &    500 &K\\
\logg   &   1 : 5   &    1.0 &dex\\
\feh    &   $-3$ : 0    &   0.50 &dex\\
\bottomrule
\end{tabular}
\label{tab:grids}
\end{table}

\section{Analysis code: FERRE}\label{sec:methods}

In \TOH and \RJ, similar codes were used in the analysis, namely spexxy\footnote{\url{https://spexxy.readthedocs.io}} and ULySS, respectively. The main difference is in the adopted models: while \TOH used the GSL synthetic library, \RJ, employed empirical spectra. We used FERRE, a public code for full-spectrum fitting, notably used in the SEGUE stellar parameter pipeline \citep{lee2008a} and in the APOGEE pipeline \cite[ASPCAP]{garciaperez2016}. Briefly, it interpolates between the nodes of an evenly sampled grid of synthetic stellar models and performs a $\chi^2$ minimisation to find the parameters of the model that best reproduce the observed spectrum. We used two different model grids: GSL and \AP (described in Sect.\,\ref{sec:anal_gsl} and \ref{sec:anal_ap}).

FERRE offers a choice of optimisation algorithms (we used the default Nelder-Mead minimisation algorithm\footnote{We checked that the other minimisation algorithms accurately produce the same results for any interpolation scheme. Therefore, we adopt this default algorithm, and will not discuss the other algorithms.}), of different interpolation schemes in the grid, and of different normalisations of the spectra to their pseudo-continuum.
Thus, it allows us to investigate how the different algorithms of interpolation in the grids could affect the measurements. 

In spexxy or ULySS, the broadening function and systemic velocity are fitted at the same time as the atmospheric parameters, as well as a multiplicative polynomial that absorbs the differences in flux calibration between the observation and the reference spectra. Fitting these parameters simultaneously with the atmospheric parameters avoids biases that may occur when the different parameters are partially degenerate, but it requires more computations. In contrast, FERRE does not adjust the broadening and continuum: the observations and the model grid should have the same LSF and be normalised in the same way. These are the essential differences between the two approaches.

\subsection{Equivalent \notep{iron abundance}}
\label{sec:equiv_feh}

While classical abundance methods at high spectral resolution (R$\sim 40\,000$) measure individual resolved lines, full-spectrum fitting uses the complete spectral information by combining all the spectral lines from any chemical species present in the considered wavelength range. This is an optimal use of the signal, and thus opens the method to the exploitation of low S/N spectra (e.g. in SEGUE, \feh was determined for spectra with S/N > 10 by \citealt{lee2008a}), and of low- and medium-resolution spectra.

A rigorous application of this method would be to simultaneously determine the abundances of all the elements that significantly contribute to the opacity in the considered wavelength range. This would be very difficult for two reasons. Firstly, because the larger the number of free parameters, the more unstable the fit. Secondly, because this will require a grid with separate dimensions sampling each chemical element. This is not practically feasible because of the excessively large volume of data it would represent, and because of the time needed to compute such a grid.
The usual simplification is to fit only one or two abundance parameters, typically \feh or \feh plus \afe, and assume a fixed (solar) pattern for the other elements: for this reason, we refer to the metallicity as the \textit{\textup{equivalent}} iron abundance (\fehe). A mismatch between the assumed and real pattern will result in biases of the derived parameters.
For example, if a spectrum enhanced in $\alpha$-elements is fit with solar-scaled reference spectra,  the measured equivalent \fehe would be larger than the true one: the iron lines will be over-fitted, and the $\alpha$-elements will be under-fitted.
A strategy for measuring the abundance of individual $\alpha$-elements has been tried at higher spectral resolution in the ASPCAP pipeline \citep{garciaperez2016}. It consists of determining \feh and \afe first, and then using spectral masks to isolate the lines of each element to measure its abundance. This approach is probably difficult to use at the MUSE resolution, where the lines are blended. 

In the following analysis, we use a fixed abundance pattern, either with $\afe\,=\,0$, or with enhanced $\alpha$-elements ($\afe\,=\,0.4$ or $0.5$ dex) corresponding to the pattern expected in NGC\,6397. The first equivalent \fehe is expected to over-estimate the real \notep{iron abundance}, and the second is expected to be a more accurate estimate.

\subsection{Interpolation schemes}

As already mentioned in the previous section, FERRE has different options for interpolating the models at any arbitrary location in the parameter space, namely, linear interpolation, and quadratic and cubic Bezi\'er splines following the description in \citet{auer2003}. These methods are local interpolations, computed on two (for linear) to four (for cubic interpolations) neighbours on each axis.\footnote{FERRE also proposes a method called ``cubic spline'', based on \citet{press92}, but while the scope of the cubic spline is to provide an interpolating function with continuous first and second derivatives over the whole range of the parameters, the implementation is ``local'' (solved in each dimension in a four-point neighbourhood).}

Although all the interpolation schemes implemented in FERRE are locally infinitely derivable over their computation interval,
the overall interpolating function is continuous but not derivable at the nodes. On each side of a node, a different interpolation neighbourhood is used, and the local interpolating functions are therefore different.
In contrast, a true spline interpolation, like in \citet{press92}, would be derivable at the nodes, but it would be more demanding in computation time and memory storage.

\citet{meszaros13b} evaluated the accuracy of the different interpolation schemes. They computed synthetic models at random locations in the parameter space between the nodes of the grids and compared them to those derived by interpolating the grid. They found that the cubic interpolation schemes produce residuals that are about 40\% smaller than the linear interpolation, and are marginally better than the quadratic interpolation. We therefore adopted the cubic Bezi\'er spline interpolation as a default choice (i.e. we use this method when not specified otherwise), and we used the other methods to perform comparisons.

\subsection{Normalisation to the pseudo-continuum}

The observations and the models were normalised in the same way before the minimisation was run. This is essential to avoid that a mismatch in the shape of the continuum does not contribute to the $\chi^2$, which would unavoidably happen even with the most carefully flux-calibrated spectra (because of uncertainties in the flux calibration or in the correction for Galactic extinction). This normalisation was made using a running average implemented in FERRE. The code also has two other normalisation methods that we did not find appropriate for the present work: polynomial, which is limited to very low degrees, and segmented average, which does not remove the local shape of the spectral energy distribution.

\subsection{Default set-up}
\label{sec:default}

Unless otherwise indicated, we adopted the following setup. 
We run FERRE on the complete MUSE wavelength range, $\lambda\lambda = 4749.52 - 9300$\,\AA. The effect of using different wavelength ranges is discussed in Sect.\,\ref{sec:discussion}.
We adopted the cubic interpolation, because this option was preferred by \citet{meszaros13b} and it indeed produces satisfactory results. We evaluate the importance of this choice in Sect.~\ref{sec:anal_gsl_interpolation}. 
As normalisation scheme, we chose the running mean with a default window of 60 pixels (i.e. 75\,\AA), as with this value the residuals (i.e. the difference between the observed and fitted spectra) appear flat. We evaluate the sensitivity of this parameter in Sect.~\ref{sec:anal_gsl_normalisation}.
As initial guesses for the stellar parameters, we set $\teff = 5500$\,K and $\feh = - 1.9$\,dex. We tested that other choices, such as starting from the photometric \teff or from $\teff = 4500$\,K, do not affect the solutions. As in \TOH and \RJ, we kept \logg fixed to the photometric value \notep{(see Sect.~\ref{sec:data_logg}).}
We fixed $\afe\,=\,0.4$\,dex (for GSL) or 0.5\,dex (for AP), as it corresponds to the abundance pattern expected in the cluster. We explore the effect of using $\afe\,=\,0$ in Sect. 5.1.

\begin{figure*}[]
\centering
\includegraphics[scale=1.00]{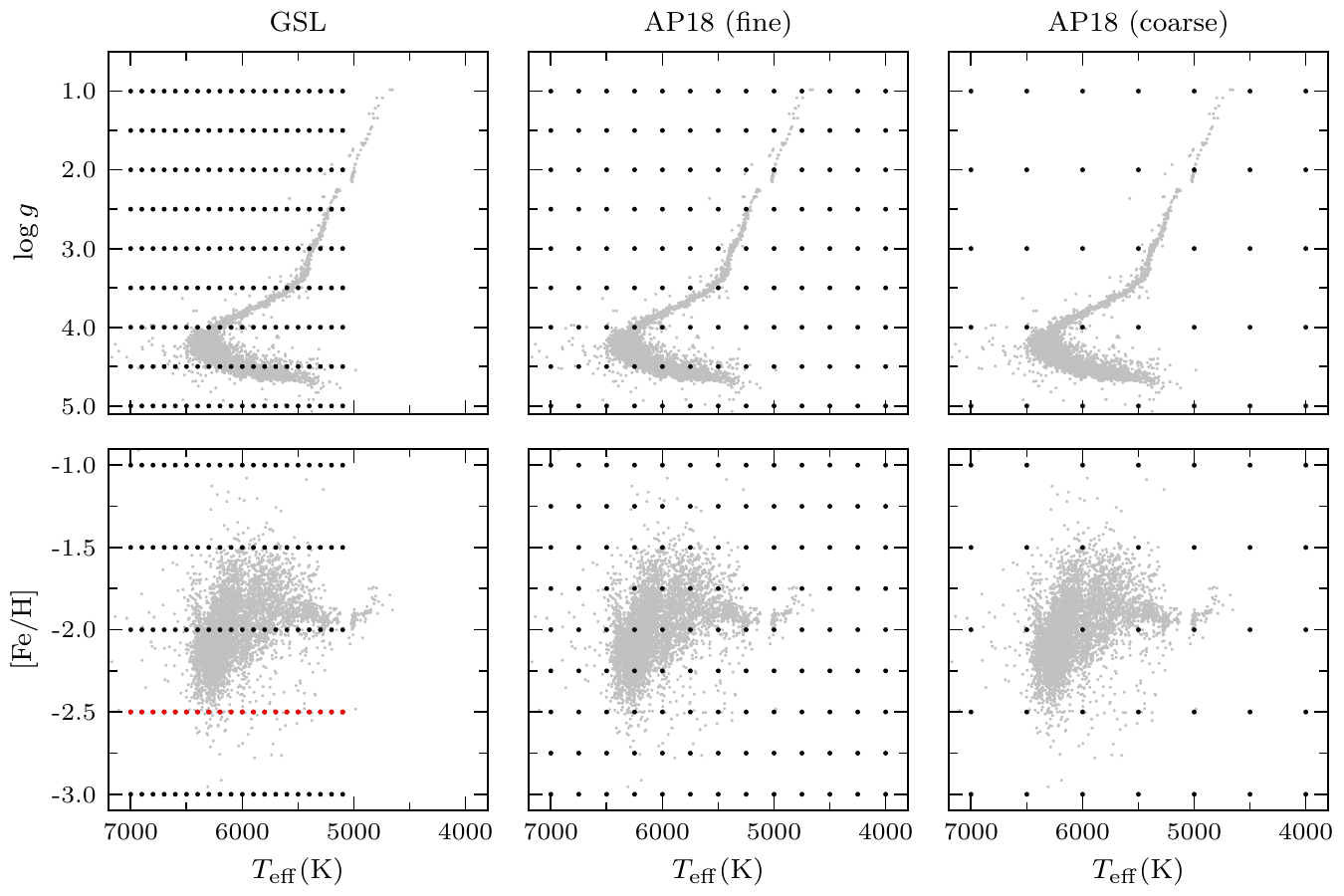}
\caption{Parameter space coverage of the grids. The grey dots represent the analysed sample, and the black dots mark the nodes of the grids of synthetic spectra. The top panels are the projections on the \logg vs. \teff plane, and the bottom panels show the projections in \feh vs. \teff. The left column is for the GSL (the $\feh = -2.5$\,dex slice, which is not originally a part of GSL, and is produced by averaging the $\feh = -3$\,dex and $\feh = -2$\,dex slices, see Sect.~\ref{sec:anal_gsl}). The central column is the \AP grid (see Sect.~\ref{sec:anal_ap}). The right column is also from the \AP grid, but with twice larger steps on each axis to evaluate the effect of a coarser grid (see Sect.~\ref{sec:anal_ap_coarsegrid}).}
\label{fig:grids}
\end{figure*}

\section{Analysis with the GSL grid}
\label{sec:anal_gsl}

GSL is a grid of high-resolution synthetic spectra created with the PHOENIX stellar atmosphere code \citep{hauschildt99}. The spectra were computed in spherical geometry (important in the low-gravity regime) and in local thermal equilibrium (LTE). The structure of the models is similar to the structure of MARCS models \citep{2008gustafsson}, and they were computed adopting the PHOENIX-ACES equation of state and opacities. The adopted line list comes from \citet{2009kurucz}. GSL covers a wide range of parameters: $2\,300<~\teff<~12\,000$\,K\footnote{The original grid was limited to $\teff \leq 8\,000$\,K. The version available on the website extends the grid to hotter models.}, $ 0<~\logg<6$, $-4 < \feh < 1$\,dex, and $-0.2<~\afe<~1.2$\,dex.

In GSL, as extensively described in \citet{2013Husser}, the micro-turbulence velocity parameter, $\xi$, is computed from the macro-turbulence velocity, that is, from the convection, and it varies from one model to the other. Its values are stored in the header of the fits files.  Abundance ratios are calculated with respect to the solar reference abundances given by \cite{2009asplund}. 

We used the version post-processed to a reduced spectral resolution, R $= 10\,000$, over the wavelength range from 3000 to 25000\,\AA.
We selected grids with fixed $\afe = 0$ and $+0.4$ dex (the latter being a reasonable assumption for low-metallicity GCs), and we restricted the range of the other parameters to what is suitable for the cluster. 
At low gravity, $\logg = 0$ and $0.5$ dex, the temperature coverage of GSL is restricted. For instance, at $\feh = -2$ dex, for $\logg = 0$ dex, the coverage is $2300 <~\teff~< 6000$\,K. The absence of hotter models is an obstacle for making a grid, because FERRE requires a strictly rectangular grid. We therefore chose to truncate the sample to $\logg > 1$.

\begin{figure*}[]
    \subfloat{\includegraphics[scale=1.00]{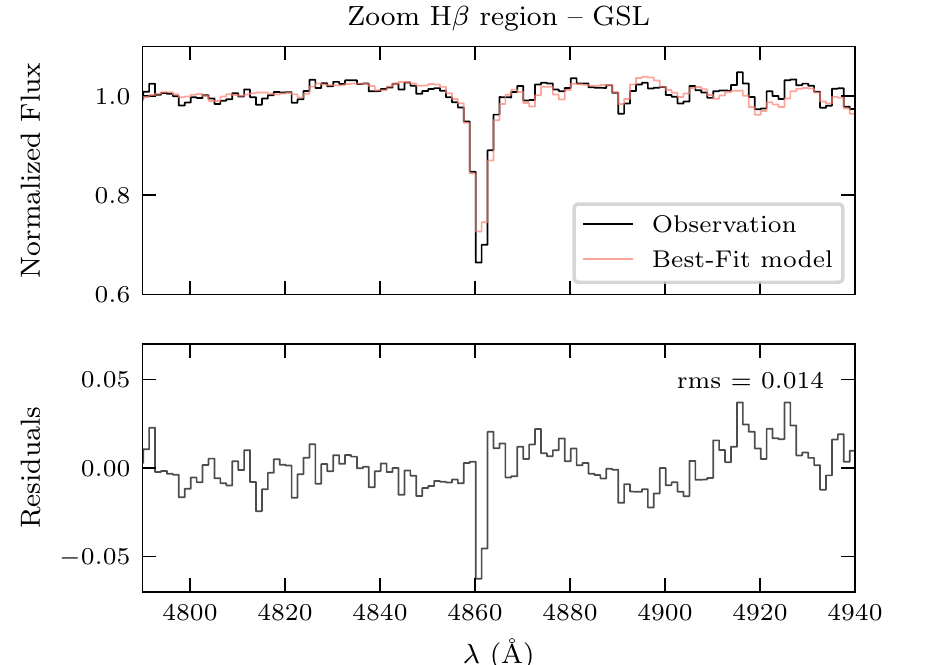}}
    \subfloat{\includegraphics[scale=1.00]{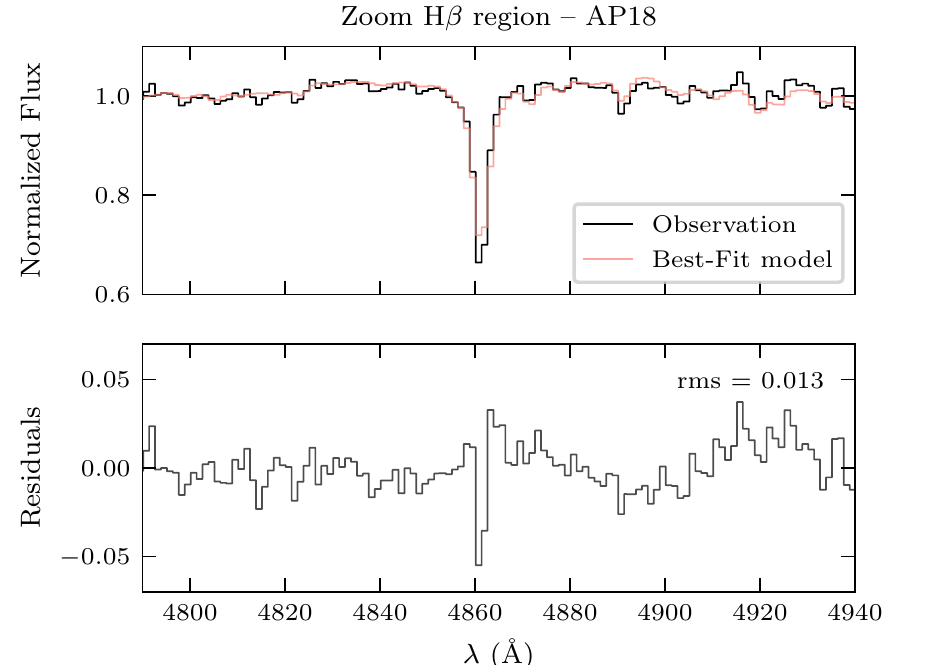}}\\
    \subfloat{\includegraphics[scale=1.00]{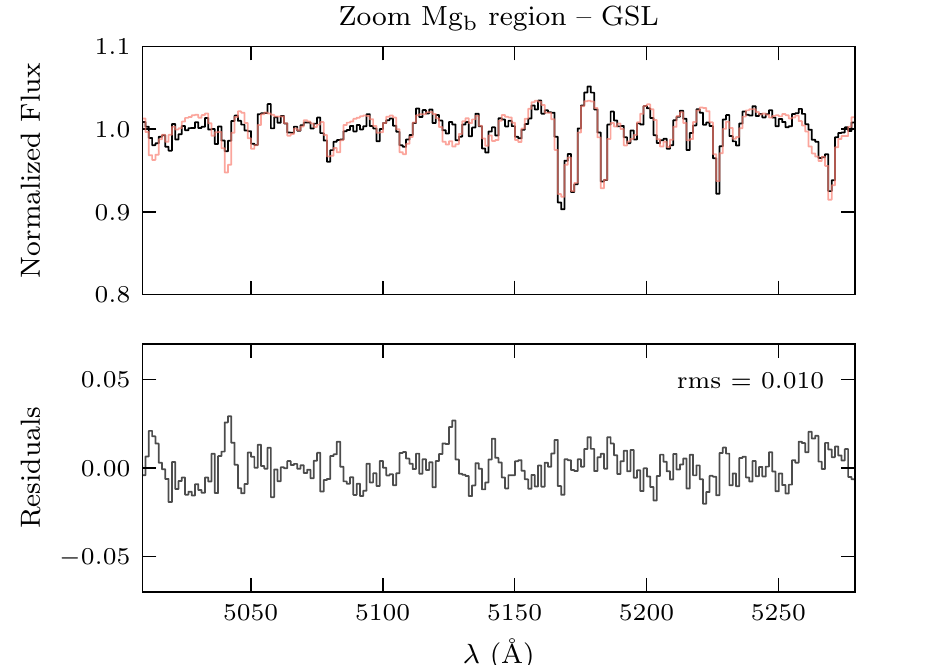}}
    \subfloat{\includegraphics[scale=1.00]{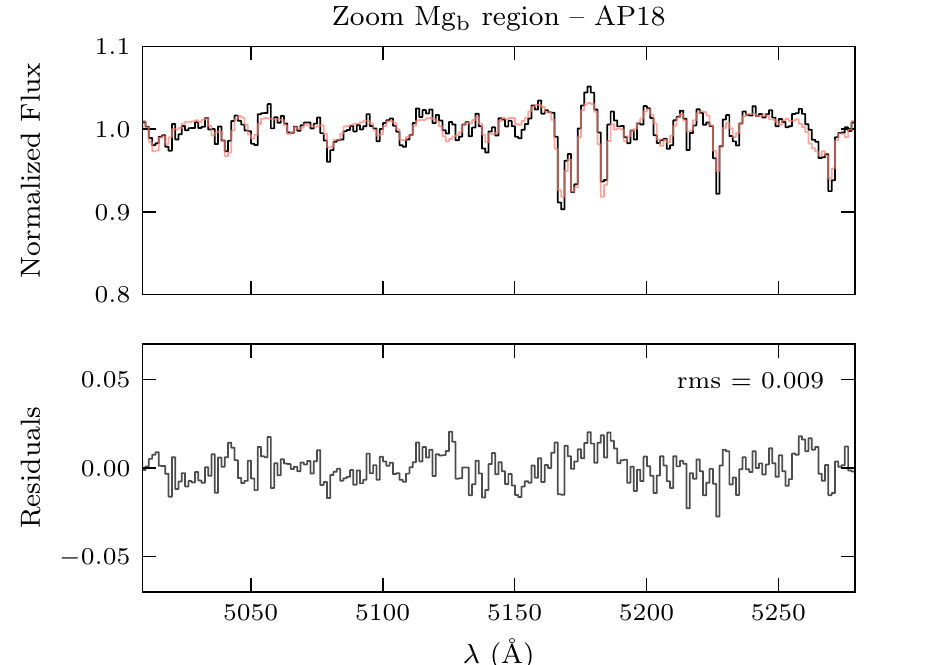}}\\
    \subfloat{\includegraphics[scale=1.00]{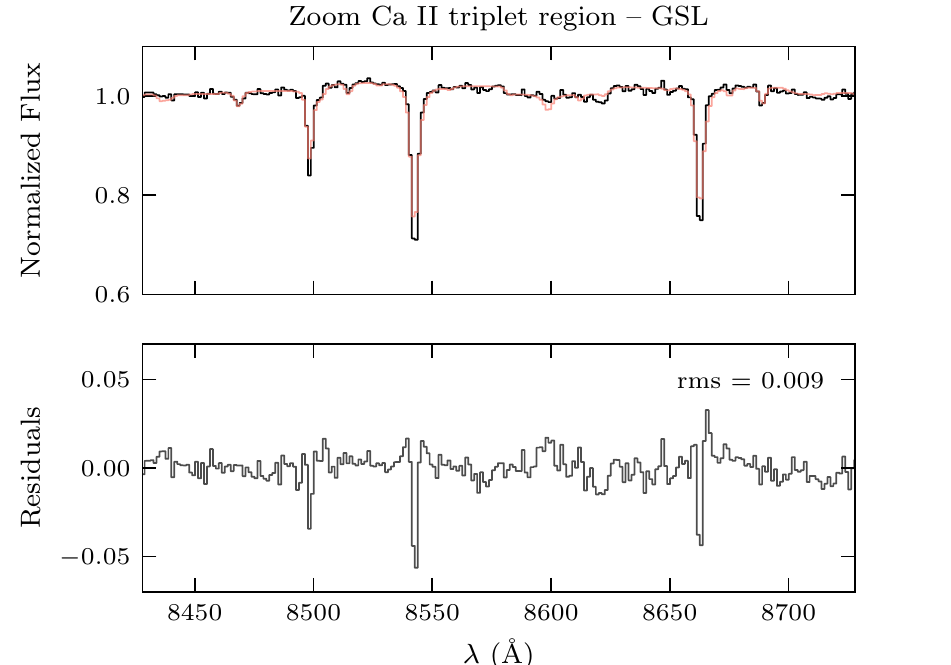}}
    \subfloat{\includegraphics[scale=1.00]{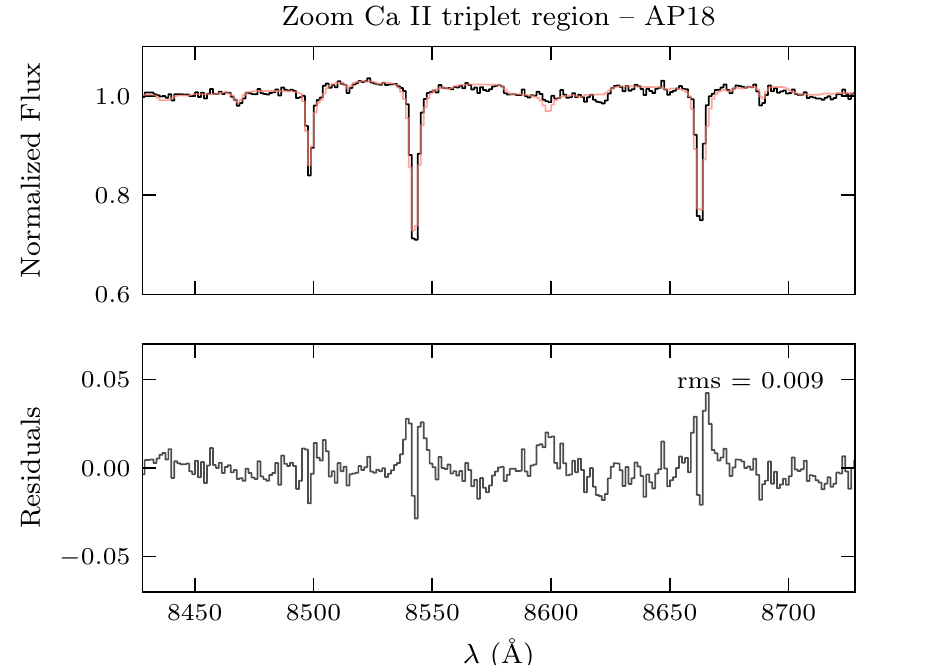}}
    \caption{
      Typical fits of a spectrum (ID = {\tt id000010747jd2456866p4985f000}). The fit, performed over the complete wavelength range, 4750 -- 9300 \AA, returned $\teff = 5471$\,K and $\fehe = -1.84$\,dex for global rms residuals of $0.01$ for both GSL and \AP. The left column shows the $\afe = 0.4$\,dex GSL grid, and the right column presents the $\afe = 0.5$\,dex fine \AP grid. The top row shows a small wavelength region around the H$_\beta$ line, the middle row shows a region around the Mg$_{\rm b}$ triplet, and the bottom one shows a region around the Ca\,\textsc{ii} triplet.
      For each plot, the top panel presents the observation (black line) with the best-fit model over-plotted in red. The bottom panel presents the residuals, observation minus model. 
    }
    \label{fig:residuals_example}
\end{figure*}

\begin{table*} 
\small{
\caption{Mean \notep{iron abundance}, errors, dispersion, and trend obtained from different setups, compared with \TOH and \RJ.}
\centering

\begin{tabular}{lccc  @{\hspace{\tabcolsep}} c c c c >{$}D..{2}<{$}}  
  
  \toprule
Lib. & \afe & $\lambda\lambda$ (nm) & I &\mfeh& \mc{$\epsilon$(\teff)}& \mc{$\epsilon$($\feh$)}& \mc{$\sigma$($\feh$)}& \mc{$\tau$($\feh$)} \\
\multicolumn{1}{c}{(1)} & \multicolumn{1}{c}{(2)} & \multicolumn{1}{c}{(3)} & \multicolumn{1}{c}{(4)} & \multicolumn{1}{c}{(5)} & \multicolumn{1}{c}{(6)} & \multicolumn{1}{c}{(7)} & \multicolumn{1}{c}{(8)} & \multicolumn{1}{c}{(9)} \\
\midrule
\multicolumn{9}{c}{\cite{husser16}}\\

GSL& 0 & $475 - 930$ & C & $-1.995$ &46& 0.080&  0.103&  $0.21$ \\
\midrule
\multicolumn{9}{c}{\cite{jain20}}\\
MIL& SN & $475 - 740$ & P & $-2.086$ &56& 0.078&  0.095&  $-0.05$ \\
\midrule\midrule

\multirow{3}{*}{GSL}& \multirow{3}{*}{$0.4$} & \multirow{3}{*}{$475 - 930$} & L & $-1.966$ &41& 0.060&  0.081&  $0.15$ \\
&  &  & Q & $-1.942$ &40& 0.063&  0.084&  $0.17$ \\
&  &  & C & $-1.945$ &41& 0.059&  0.082&  $0.18$ \\

\hline
\multirow{3}{*}{GSL}& \multirow{3}{*}{$0.4$} & \multirow{3}{*}{$475 - 740$} & L & $-2.186$ &50& 0.131&  0.137&  $0.28$ \\
&  &  & Q & $-2.180$ &50& 0.134&  0.142&  $0.28$ \\
&  &  & C & $-2.167$ &51& 0.133&  0.139&  $0.25$ \\

\hline

\multirow{3}{*}{GSL}& \multirow{3}{*}{$0.4$} & \multirow{3}{*}{$475 - 580$} & L & $-2.224$ &56& 0.146&  0.159&  $0.33$ \\
&  &  & Q & $-2.218$ &55& 0.147&  0.164&  $0.33$ \\
&  &  & C & $-2.203$ &57& 0.149&  0.165&  $0.31$ \\

\midrule\midrule

\multirow{3}{*}{GSL}& \multirow{3}{*}{$0$} & \multirow{3}{*}{$475 - 930$} & L & $-1.727$ &42& 0.063&  0.086&  $0.19$ \\
&  &  & Q & $-1.704$ &42& 0.061&  0.082&  $0.16$ \\
&  &  & C & $-1.712$ &42& 0.061&  0.082&  $0.16$ \\

\hline
\multirow{3}{*}{GSL}& \multirow{3}{*}{$0$} & \multirow{3}{*}{$475 - 740$} & L & $-2.066$ &48& 0.130&  0.135&  $0.22$ \\
&  &  & Q & $-2.055$ &48& 0.133&  0.143&  $0.23$ \\
&  &  & C & $-2.047$ &48& 0.125&  0.131&  $0.21$ \\

\hline

\multirow{3}{*}{GSL}& \multirow{3}{*}{$0$} & \multirow{3}{*}{$475 - 580$} & L & $-2.129$ &56& 0.151&  0.168&  $0.34$ \\
&  &  & Q & $-2.123$ &56& 0.154&  0.174&  $0.34$ \\
&  &  & C & $-2.109$ &56& 0.154&  0.171&  $0.33$ \\

\midrule\midrule

\multirow{3}{*}{\AP}& \multirow{3}{*}{$0.5$} & \multirow{3}{*}{$475 - 930$} & L & $-2.031$ &41& 0.055&  0.079&  $0.20$ \\
&  &  & Q & $-2.015$ &40& 0.052&  0.075&  $0.21$ \\
&  &  & C & $-2.014$ &41& 0.052&  0.075&  $0.21$ \\

\hline
\multirow{3}{*}{\AP}& \multirow{3}{*}{$0.5$} & \multirow{3}{*}{$475 - 740$} & L & $-1.945$ &48& 0.086&  0.113&  $0.46$ \\
&  &  & Q & $-1.928$ &47& 0.086&  0.112&  $0.46$ \\
&  &  & C & $-1.925$ &47& 0.084&  0.111&  $0.46$ \\

\hline

\multirow{3}{*}{\AP}& \multirow{3}{*}{$0.5$} & \multirow{3}{*}{$475 - 580$} & L & $-2.092$ &61& 0.129&  0.137&  $0.39$ \\
&  &  & Q & $-2.074$ &62& 0.128&  0.134&  $0.40$ \\
&  &  & C & $-2.073$ &62& 0.127&  0.134&  $0.41$ \\

\midrule\midrule

\multirow{3}{*}{\AP}& \multirow{3}{*}{$0$} & \multirow{3}{*}{$475 - 930$} & L & $-1.650$ &42& 0.058&  0.081&  $0.24$ \\
&  &  & Q & $-1.636$ &41& 0.056&  0.079&  $0.25$ \\
&  &  & C & $-1.635$ &42& 0.056&  0.079&  $0.25$ \\

\hline
\multirow{3}{*}{\AP}& \multirow{3}{*}{$0$} & \multirow{3}{*}{$475 - 740$} & L & $-1.758$ &50& 0.089&  0.120&  $0.47$ \\
&  &  & Q & $-1.744$ &50& 0.091&  0.122&  $0.47$ \\
&  &  & C & $-1.741$ &50& 0.091&  0.121&  $0.47$ \\

\hline
\multirow{3}{*}{\AP}& \multirow{3}{*}{$0$} & \multirow{3}{*}{$475 - 580$} & L & $-1.925$ &62& 0.145&  0.148&  $0.42$ \\
&  &  & Q & $-1.908$ &63& 0.143&  0.147&  $0.44$ \\
&  &  & C & $-1.907$ &63& 0.143&  0.146&  $0.45$ \\

\midrule\midrule

\multirow{3}{*}{APC}& \multirow{3}{*}{$0.5$} & \multirow{3}{*}{$475 - 930$} & L & $-2.068$ &42& 0.061&  0.083&  $0.20$ \\
&  &  & Q & $-2.023$ &41& 0.055&  0.077&  $0.20$ \\
&  &  & C & $-2.013$ &40& 0.051&  0.074&  $0.22$ \\

\bottomrule
\end{tabular}
\tablefoot{(1) Grid or library; MIL stands for the MILES library \citep{sanchez-blazquez2006} used in \RJ, and APC for the \AP coarse grid. (2) \afe value; SN stands for solar neighbourhood, as the abundance pattern of the empirical library used by \RJ is made of stars from this environment. (3) Wavelength range in nm. (4) Interpolation schema; L: linear, Q: quadratic, C: cubic, P: polynomial approximation; the cubic splines used in \TOH are different from the cubic Bezi\'er interpolation of FERRE used in our analyses.  (5) \mfeh is the average of the mean \notep{iron abundance} determined in five \teff bins of 250 K, for $5100 < \teff < 6350$\,K. (6) $\epsilon(\teff)$ is the mean uncertainty on \teff estimated from pairs of repeated observations, averaged in the five \teff bins. (7) $\epsilon(\feh$) is the mean uncertainty on $\fehe$ estimated from pairs of repeated observations and computed in the same way as $\epsilon(\teff$). (8) $\sigma(\feh)$ is the square root of the mean variance of $\fehe$ in the same five bins.
  (9) $\tau(\feh)$ is the equivalent \notep{iron abundance} trend, computed as the difference of mean \fehe in the first and last bin. 
}

\label{tab:feh_spread}
}
\end{table*}

\subsection{Construction of the grids and analysis}
\label{sec:fits}

In figure\,8 of \TOH, we noticed an irregular distribution of the measurements along the \teff sequence. 
We investigate these issues in Appendix\,\ref{sec:facitsaltum} and find a non-physical discontinuity at $\teff = 5000$\,K.  For this reason, we limited the range of our grid to $\teff~\geq~5100$\,K. 
The characteristics of the resulting grid are presented in Table~\ref{tab:grids}.
FERRE requires the grid to sample each axis evenly\footnote{This aspect, not fundamental to the FERRE principle, is due to the choice of describing the grid with a simple coordinate system.}, but the GSL \feh sampling is not uniform. We show in Fig.~\ref{fig:grids} (left panel)  the parameter space coverage of the grids (the top and bottom rows show the \logg versus \teff and \feh versus \teff planes, respectively), we added the missing $\feh = -2.5$\,dex slice  to GSL by averaging the $-2.0$ and $-3.0$\,dex slices (red dots in Fig.~\ref{fig:grids}). 
In Appendix\,\ref{sec:facitsaltum}, we also found an outlier near the TO, and we replaced it with the average between the spectra 100\,K apart in the grid. This outlier was found to affect the measurements, while the two others detected in this region were not found to have an impact.

The procedure of building the grid in the FERRE format has three steps. i) Conversion from vacuum into air wavelengths,
ii) injection of the relative LSF as described in Sect.~\ref{sec:data}, and
iii) rebinning as the observations.

When the grid was ready, we ran FERRE. Figure~\ref{fig:residuals_example} (left column) presents the typical fit of a stellar spectrum. It shows three remarkable regions: around H$_{\beta}$ (spectral window 4790-4940\,\AA), Mg$_{\rm{b}}$ (spectral window 5000-5300\,\AA), and the Ca\,\textsc{ii} triplet (spectral window 8400-8750\,\AA). The rms of the residuals for the full wavelength range (not shown in the plot) is 0.01. The test spectrum has an estimated S/N of 115, and the residuals indicate a normalised $\chi^2$ of 1.7. The figure shows that the LSF and the wavelength calibration are adequately matched, but it also shows significant residuals on strong lines. As already noted in literature, strong lines such as the H lines or the Ca\,\textsc{ii} triplet are difficult to model for a number of reasons, for instance the LTE approximation \citep[see e.g.  ][]{2007martins}.

\begin{figure*}[h!]
     \centering
    \subfloat{\includegraphics[scale=1.00]{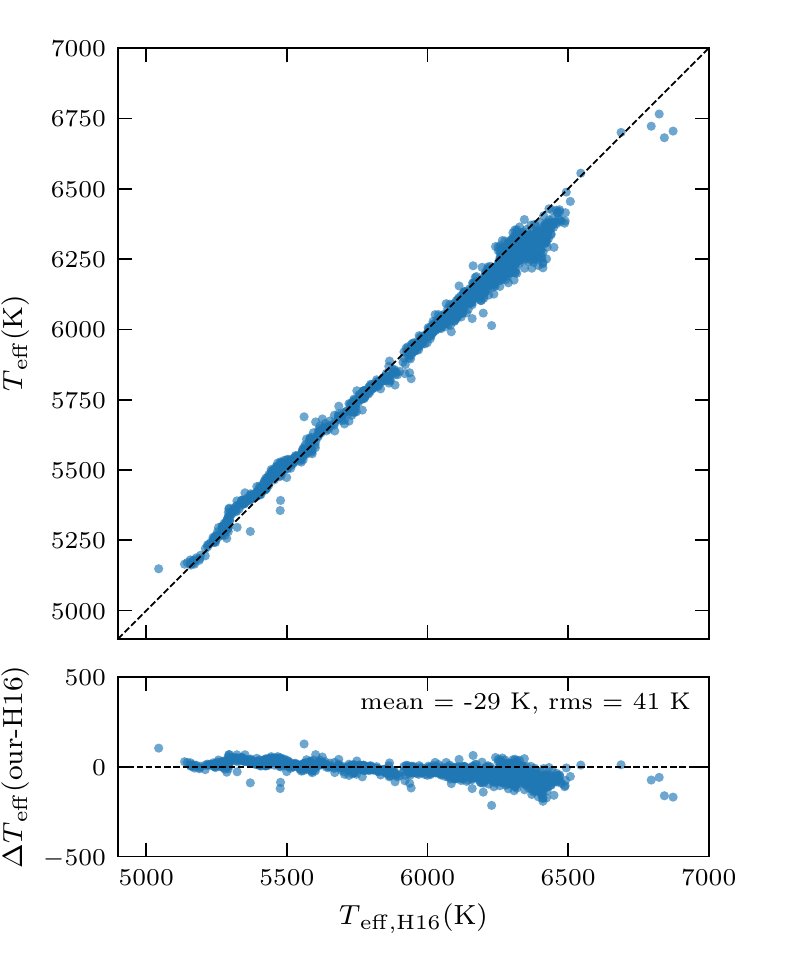}}
    \quad
    \subfloat{\includegraphics[scale=1.00]{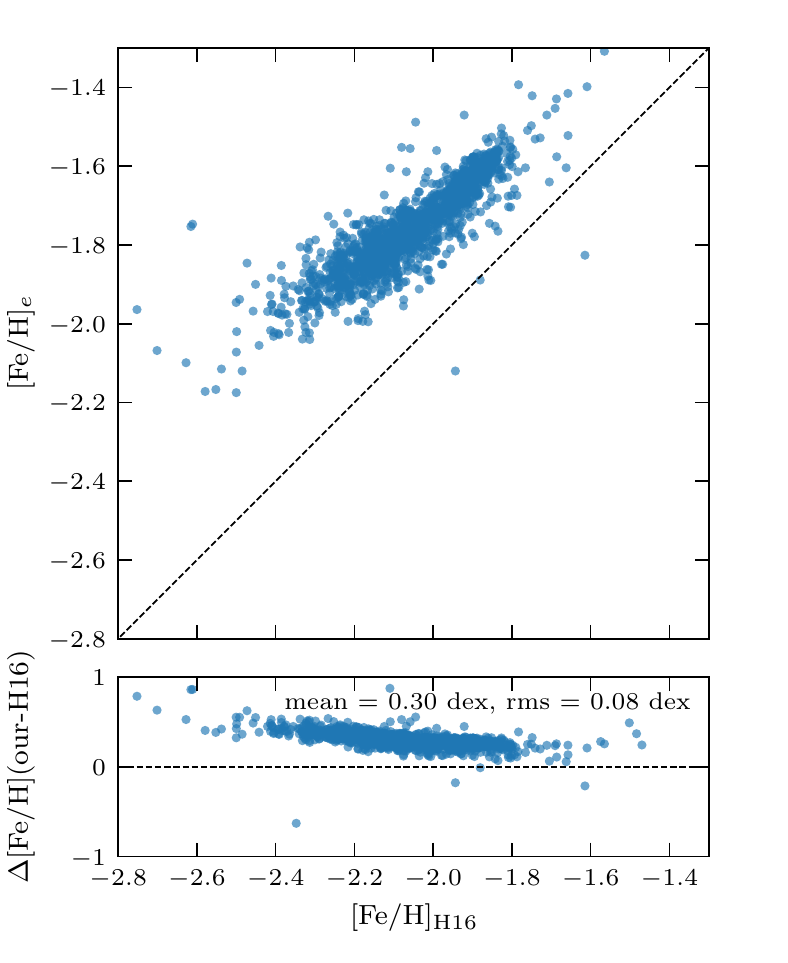}}
    \caption{Comparison between our measurements (y-axis) and \TOH results (x-axis) obtained with solar \afe. The bottom panels show the differences $\Delta$ (our $-$ \TOH). We report the comparison between the \teff on the left and the comparison between our \fehe and \feh of \TOH\ on the right. The mean and rms differences are indicated.
    }
        \label{fig:comparisons}
\end{figure*}

Table\,\ref{tab:feh_spread} reports the results obtained for the different setups we used. Our measurements of \teff and \fehe for the individual stars for each combination of wavelength, interpolation, and assumed \afe are available in electronic form at the CDS. In order to assess the precision and to compare these solutions, we estimated (i) the measurement errors and (ii) the \notep{iron abundance} spread of the sequence.
The measurement errors (due to the noise in the observations, the data reduction, and the analysis) can be reliably estimated using pairs of repeated observations. As in \RJ, given two observations (1 and 2) of the same star \textit{i}, we computed $\Delta P_{i}= (P_{1,i}-P_{2,i})/\sqrt{2}$, where $P_i$ is any of the stellar parameters (\fehe or \teff), and the measurement error $\epsilon(P_i)$ is estimated as the rms of $\Delta P_{i}$.
Because the \notep{iron abundance} may change along the sequence of the cluster, the \fehe dispersion measured over the entire sample would combine the effects of the trend and of the spread. To isolate the spread, we divided the \teff range into bins of 250\,K and computed the \fehe dispersion, $\sigma_j(\feh),$ in each bin.
For consistency, the final errors and spread,  $\epsilon(\teff)$, $\epsilon(\feh)$ and $\sigma(\feh)$, are the quadratic average of the  $\epsilon_j(P_i)$ and $\sigma_j(P_i)$ (columns 6, 7, and 8), and the mean \fehe (column 5) is the average of the mean \fehe in each bin. 
As expected, we note that $\epsilon(\feh)$ is always smaller than $\sigma(\feh)$. This reflects the fact that in addition to the effect of the noise and observation, $\sigma(\feh)$ includes other sources of errors (e.g. due to the precision of the spectral models) and a possible cosmic spread.
Finally, the equivalent iron abundance trend $\tau(\feh)$ (column 9) is defined as the difference between the mean \fehe in the first and last bins.

\subsection{Comparison with \TOH}

Figure~\ref{fig:comparisons} compares our measurements with those of \TOH, both obtained with \afe = 0. On the left, we compare our \teff (y-axis) and \teff from \TOH (x-axis) in the top panel, and at the bottom, we report the difference between our \teff and \TOH ($\Delta$\teff(our-H16)) as a function of the \teff from \TOH. On the right, we show the same set of plots for the \fehe measurements.
Our average equivalent \notep{iron abundance} is higher by 0.30\,dex than the one reported in \TOH. The rms \fehe dispersion between the two series is 0.08\,dex, which is consistent with the measurement errors. \notep{In contrast, there is no overall bias in \teff}: the mean difference is $-29$\,K with an rms of 41\,K, \notep{but there is a significant drift with a $\sim 100$\,K amplitude; our values are cooler than those of \TOH at the TO}.

The two analyses use the same data, the same reference synthetic spectra, and the same technique \notep{(full-spectrum fitting)}. The source of the \fehe bias \notep{and of the temperature drift} was searched for in the several differences between the two approaches.
The differences due to the analysis programs are that (i) we derived only the atmospheric parameters, while \TOH also adjusted the broadening and shift of the spectra, (ii) we used various schema implemented in FERRE to interpolate over the grid of models, while \TOH used a cubic spline interpolation, and (iii) we normalised the observed and model spectra to their pseudo-continuum before fitting them, while \TOH fitted a multiplicative polynomial at the same time as all the other parameters.

The significance of these three aspects is investigated in detail in the next sections: apparently, they cannot explain the difference of \fehe between the two analyses. After excluding differences in the programs as a source for the bias, a possible explanation could be searched for in the weighting schema of each individual wavelength bin.
\notep{To support this hypothesis, we note that the \TOH solution is intermediate between our solutions with the full and trimmed to $740$ nm wavelength ranges. By decreasing the weight of the near-IR (NIR) region, we can indeed recover a mean \fehe consistent with \TOH, but with larger errors. When we use a constant weight throughout the wavelength range, the \teff drift with respect to \TOH is cancelled (but the abundance difference remains). We cannot attempt to precisely reproduce the \TOH results because we lack information about the weighting that was used. This sensitivity to the weighting reflects inconsistencies in the modelling in different spectral regions, reminiscent of those emphasised by \citet{2021lancon}.}

\subsection{Effect of the LSF matching}
\label{sec:lsf_effect}

An accurate LSF injection is a priori more critical here than in \TOH and \RJ, where the velocity broadening adjusted to each individual spectrum was absorbing part of its variation throughout the MUSE sample.
We carried out two tests to evaluate the effect of the LSF on the parameters. The first test simulated the variation in resolution amongst the MUSE spectrographs, and the second test simulated wavelength shifts that could result from errors in the dispersion relation, or in the rest-frame reduction.

For the first test, we used a GSL grid with an LSF FWHM increased by 0.1\,\AA{}, equivalent to the real uncertainty.
We find that the \teff increases by 3\,K, and \fehe by $0.01$\,dex for the whole sample. This is about 10 to 20\% of the measurement errors, and therefore a minor contribution to the error budget. 

For the second test, we shifted the LSF by 0.5\,\AA{} (i.e. 0.4 times the wavelength bin). This corresponds to about 30 \kms, that is, about five times higher than the velocity dispersion measured by \TOH. Here, we find an increase in \teff by 21\,K and a decrease in \fehe by $0.03$\,dex. Therefore, the wavelength calibration uncertainty and internal velocity dispersion are also minor contributions to the total error.

We note that this second test also mimics the effect of inaccurate line lists that manifests as spectral lines modelled at incorrect wavelengths, or with incorrect strengths. This well-known problem is probably responsible for a significant fraction of the spectral mismatch between observation and models \citep{franchini18,martins19}.
By splitting the sample into two temperature regimes, we find that \fehe decreases by 0.04\,dex for the warm stars ($\teff > 5800$\,K), while it decreases by only $0.01$\,dex for the cool stars. This suggests that inaccurate line lists may affect the measured equivalent \fehe trend.

\begin{table}[!htb]
\tiny{
\caption{\fehe dispersion: Effect of the normalisation window.}
\centering
\begin{tabular}{ccccc}
\toprule
Norm. window & Interpolation & \mfeh & $\sigma$(\feh) \\
\multicolumn{1}{c}{(1)} & \multicolumn{1}{c}{(2)} & \multicolumn{1}{c}{(3)} & \multicolumn{1}{c}{(4)} \\
\midrule

\multirow{3}{*}{30}
& L & -1.950 & 0.084\\
& Q & -1.926 & 0.087\\
& C & -1.930 & 0.085\\

\hline
\multirow{3}{*}{60}
& L & -1.966 & 0.081\\
& Q & -1.942 & 0.084\\
& C & -1.945 & 0.082\\

\hline
\multirow{3}{*}{120}
& L & -1.934 & 0.086\\
& Q & -1.907 & 0.088\\
& C & -1.911 & 0.089\\

\bottomrule
\end{tabular}
\tablefoot{Dispersion along the giant branch with different normalisation windows, using GSL in the wavelength range 475 - 930 nm. (1) Running-mean normalisation window given as the number of wavelength bins. (2) Interpolation schema; L: linear, Q: quadratic, C: cubic (3) \mfeh is the average of the average \fehe determined in \teff bins of 250 K. (4) $\sigma(\feh)$ is the square root of the mean variance of \fehe determined in \teff bins of 250 K.}
\label{tab:GSL_2}
}
\end{table}

\begin{figure*}[!htb]
     \centering
      \subfloat{\includegraphics[scale=1.00]{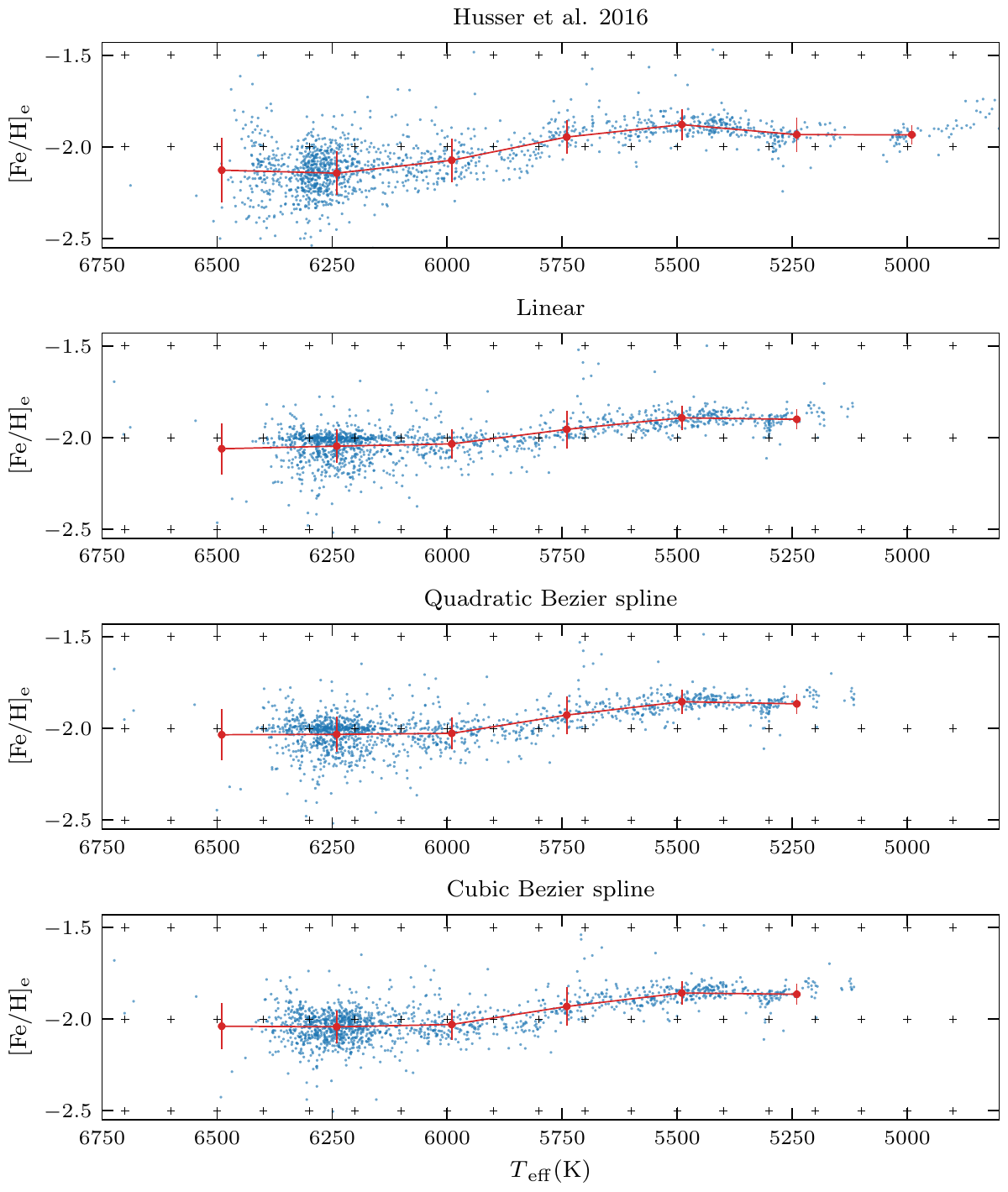}}
      \caption{\fehe vs. \teff diagrams published in \TOH (top panel, \notep{with \afe=0}) and those obtained with FERRE using GSL and the three different interpolation schemes (other panels, \notep{with \afe=+0.4}). The measurements of individual spectra are shown as blue dots, and the nodes of the grids are overlaid as thin crosses. The red dots and line are the mean \fehe in bins of 250\,K, and the error bars are the dispersion in these bins. 
}
        \label{fig:relation}
\end{figure*}

\subsection{Effect of the interpolation schema}
\label{sec:anal_gsl_interpolation}

The possible importance of interpolation has been pointed out by \RJ, who noted that the \notep{iron abundance} trend along the SG branch is only a small fraction of the mesh size. This trend is presented in Fig.~\ref{fig:relation}, in which we compare the \TOH measurements with those obtained with FERRE using various interpolation schemes, namely, linear (second panel), quadratic Bezi\'er spline (third panel), and cubic Bezi\'er spline (fourth panel). It is evident that the trend is similar with the different interpolation schemas. 
The measurement errors ($\epsilon(\feh)$ and $\epsilon(\teff)$), the mean and spread of equivalent \notep{iron abundance}, \mfeh and $\sigma(\feh)$ obtained in each case are comparable (see Table~\ref{tab:feh_spread}). This implies that the choice of the interpolation scheme is not critical in this analysis. \\

A peculiar pattern in Fig.~\ref{fig:relation} is to be mentioned: there is a concentration of measurements aligned on \fehe$ = -2.0$\,dex with the linear, and to a lesser extent with the quadratic interpolation. This is an artefact of the interpolation that is discussed in Appendix~\ref{sec:interpolation_artefact}.

\subsection{Effect of the continuum normalisation}
\label{sec:anal_gsl_normalisation}

The observations and the models were both normalised by dividing them by a running mean with a window of 60 wavelength bins, which corresponds to 75\,\AA. With this width, the residuals of the fits (Fig.~\ref{fig:residuals_example}) do not show low frequencies. We also tested different normalisation windows by fixing the width to twice larger (120) and twice smaller (30).
In Table~\ref{tab:GSL_2} we report the dispersion values corresponding to these tests. The trend is similar in the three cases, and the smaller dispersions obtained with 60 bins justify our choice.

\begin{figure}[!htb]
     \centering
      \subfloat{\includegraphics[scale=1.00]{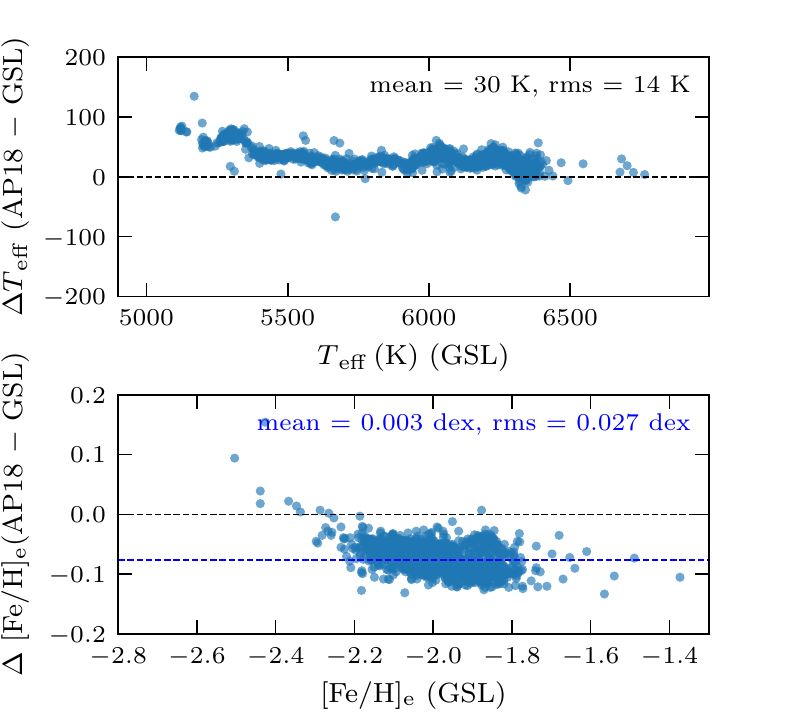}}
      \caption{Comparison between \teff and \fehe obtained with GSL and \AP grids. The top panel shows the difference between \teff ($\Delta \teff$(\AP-GSL), computed by subtracting the GSL results from the \AP results) as a function of \teff values with GSL. The bottom panel shows the difference of \notep{equivalent iron abundance} ($\Delta \fehe$) as a function of the \notep{iron abundance} with GSL.
The dashed blue line shows the expected bias due to the different level of $\alpha$-element enhancement in the two grids (0.4 vs. 0.5 dex, see the text for details). In both panels we report the mean and rms of the differences.
      }
        \label{fig:compa_gsl_ap}
\end{figure}

\begin{figure*}[!htb]
     \centering
      \subfloat{\includegraphics[scale=1.00]{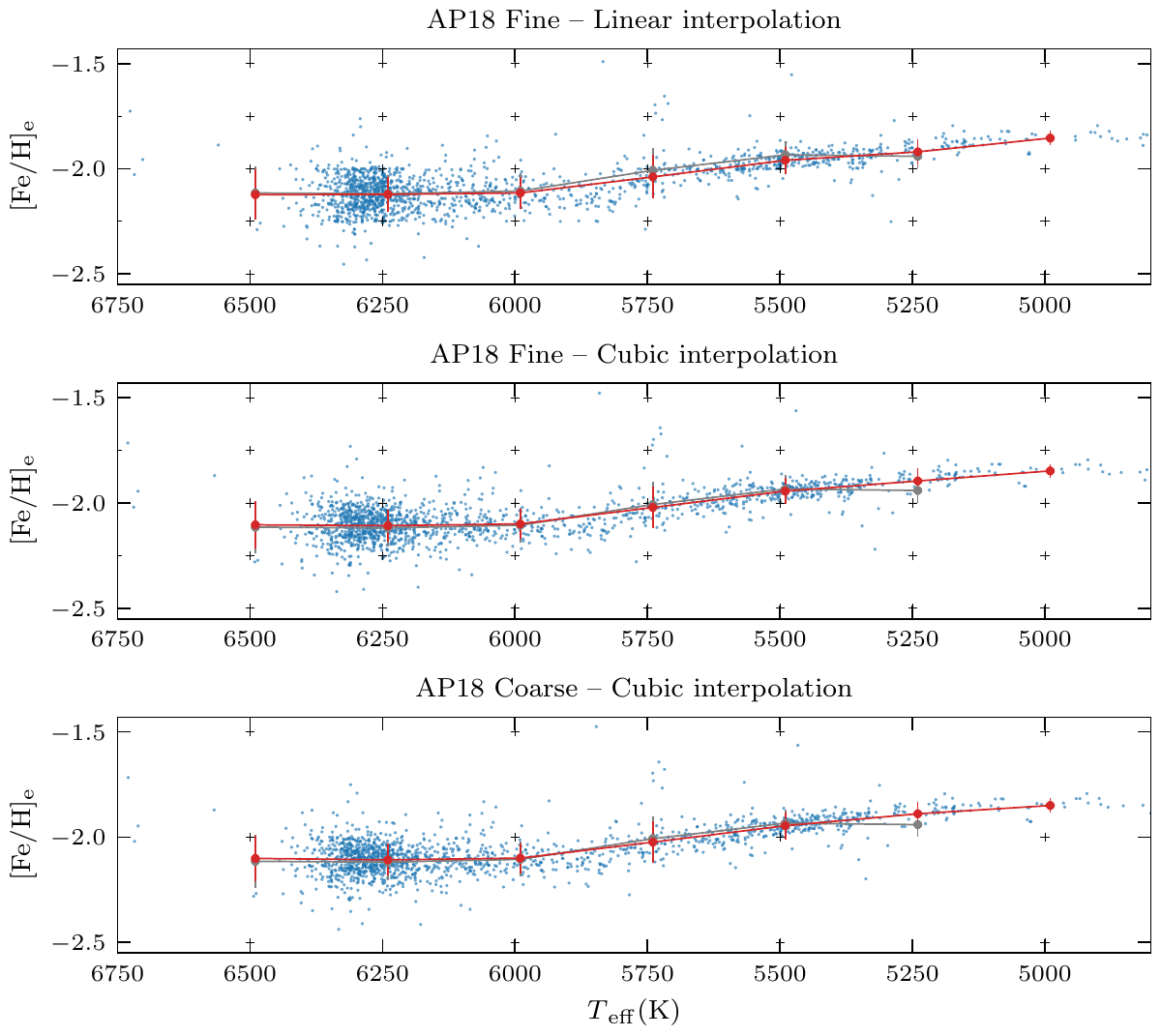}}
      \caption{\fehe vs. \teff relation using the \AP fine (top and second panels) and coarse (third panel) grids. The top panel shows the linear interpolation, and the two other panels show the cubic interpolation. \notep{The grey lines and symbols represent the solution (with an offset of -0.076) obtained with the GSL grid (cubic interpolation) and reported on Fig.~\ref{fig:relation}.}
        The \notep{other} symbols and line styles are as in Fig.~\ref{fig:relation}.}
        \label{fig:relation_ap}
\end{figure*}

\section{Analysis with \AP grid}
\label{sec:anal_ap}

\citet{allende-prieto18} provided a comprehensive set of synthetic spectral grids obtained with ATLAS9 atmosphere models, computed by \citet{meszaros12} using opacity distribution functions assuming plane-parallel and LTE approximations. The \notep{spectra} were synthesised with the radiative code ASS$\epsilon$ET \citep{koesterke2009}. \notep{The line lists are from the Kurucz public database (updated in 2007), plus some specific updates}. The grids cover the parameter space from $3500$ to $30\,000$\,K, $0$ to $5$\,dex, and $-5.0$ to $0.5$\,dex for \teff, \logg, and \feh, respectively, and also sample \afe and $\xi$.
The adopted solar reference abundance is from \citet{2009asplund}, as for the GSL grid. 
Different grids are provided at various spectral resolutions 
and with various coverage and sampling of this wide domain of parameters.

We used their so-called fine grids ns1 and ns2, at the resolution R $= 10\,000$ \citep[see table~1 from][]{allende-prieto18}, with both solar and enhanced \afe (+0.50\,dex), and with $\xi=1$ and 2\,\kms. From our fine grid (\AP in Table\,\ref{tab:feh_spread}), we extracted a coarse grid (APC in Table\,\ref{tab:feh_spread}) with a twice larger mesh size. Details can be found in Table\,\ref{tab:grids} and in Fig.~\ref{fig:grids}.

To build our final \AP and APC grids, we processed them as done for the GSL grid, that is, with the same procedure of LSF injection and rebinning. The only difference was that the \AP grids were already in air wavelengths. 

\subsection{Comparison with our GSL results}
\label{sec:compa_gsl}

We repeated the fits of the previous section with the \AP fine-grained grid with $\afe = 0.5$\,dex and $\xi = 1$\,\kms for the three interpolation schemes.
A typical fit is presented in Fig.~\ref{fig:residuals_example} and can be compared with the fit of the same spectrum with GSL made in Sect.~\ref{sec:anal_gsl}. A quantitative comparison of spectral fits with GSL and \AP grids is made by measuring the mean rms residuals over the whole sample in different spectral regions. We find that these values in the H$_{\beta}$, Mg$_{\rm b}$ , and Ca\,\textsc{ii} triplet regions are almost the same with both grids.
In the full wavelength range (not shown in the plot), as seen with the GSL grid, the rms residual is equal to 0.01.

A comparison of the solutions with \AP and GSL grids and cubic interpolation is presented in Fig.~\ref{fig:compa_gsl_ap}.
A systematic difference of equivalent \fehe is expected because the two grids are computed at different $\alpha$-elements enhancements, 0.5\,dex for \AP, and 0.4\,dex for GSL (see Sect.~\ref{sec:equiv_feh}). The solutions with solar-scaled and enhanced grids are reported in Table~\ref{tab:feh_spread}. For solar-scaled models, when the full wavelength range is considered, the equivalent \fehe are increased by 0.233 and 0.379 dex for GSL and \AP, respectively. This important effect shows that the $\alpha$-elements are a major contribution to the total opacity in the considered spectral region. By extrapolating this effect linearly, we can correct the solutions obtained with the two grids to a mid-point enhancement of $0.45$ dex. For GSL, the correction is $-0.029$\,dex, and for \AP, it is $0.047$\,dex. Hence, the mean difference is $\Delta\fehe({\rm AP18-GSL}) = -0.076$\,dex. This is represented in the figure (dashed blue line), and it closely matches the observed bias of $-0.073$ dex.

The mean biases (AP18 $-$ GSL, reported in the figure) are well within the generally accepted external errors.
The rms dispersions between the two series are two to three times smaller than the uncertainties inferred from pair-wise comparisons. This indicates that in our dataset, the random errors linked to the choice of synthetic grid contribute marginally to the error budget.

The \fehe distributions along the \teff sequence using either linear or cubic interpolation are represented in Fig.~\ref{fig:relation_ap} (top and central panels). The equivalent \notep{iron abundance} trend is qualitatively similar to the trend obtained before with GSL: this is also corroborated by the values of $\tau$(\feh) reported in Table\,\ref{tab:feh_spread}. As with GSL, the choice of the interpolation scheme does not affect the trend.   

Table\,\ref{tab:feh_spread} also shows that the estimated errors ($\epsilon$) and the spreads ($\sigma$) are slightly smaller with \AP than with GSL (by roughly 0.01--0.02\,dex).
This is consistent with \citet{franchini18}, who also compared GSL and other grids of synthetic spectra to high-resolution spectra (R $\sim$ 50\,000) obtained in the \textit{Gaia}-ESO survey \citep{2012gilmore} over a $500$\,\AA{} spectral range around $5200$\,\AA. They found that GSL matches the spectra less well than other libraries.

\subsection{Effects of different mesh sizes}
\label{sec:anal_ap_coarsegrid}

A valuable characteristic of the \AP grid is its fine \feh sampling. To test the importance of the mesh size, we used the coarse grid (see Table~\ref{tab:grids}), whose \feh sampling is the same as that of \TOH. The \fehe mean value and spread for the coarse grid with different interpolation schema are given in Table~\ref{tab:feh_spread}, and the \fehe versus \teff diagram with cubic interpolation is shown in Fig.~\ref{fig:relation_ap} (bottom panel).
The values of $\epsilon$($\feh$) with the coarse grid are not significantly higher than those with the fine grid, and the mean \notep{iron abundance} is not affected when quadratic or cubic interpolation is used. With linear interpolation, the dispersion is marginally increased and the mean \notep{iron abundance} is marginally decreased.
We conclude that the mesh size is not critical, and that a $0.5$\,dex \notep{iron abundance} step does not affect the results dramatically even for effects ten times smaller than the mesh.
Together with the minor effect of changing the interpolation schema seen in Sect.~\ref{sec:anal_gsl_interpolation}, this shows that (i) a linear approximation within the considered grids of models is usually satisfactory, and (ii) using a higher-order interpolation schema with a grid mesh as large as $0.5$\,dex in \feh allows us to recover the precision of a twice finer mesh.

\subsection{Effects of micro-turbulence velocity}
\label{sec: vturb_effects}

Micro-turbulence velocity ($\xi$) is an ad hoc parameter introduced to account for the difference between the measured and expected strength of spectral lines. The $\xi$ parameter works as an additional broadening and can be derived spectroscopically by imposing that weak and strong iron lines match with the same abundance (see \citealt{2011mucciarelli} for an extensive discussion). It also depends on the other atmospheric parameters: $\xi$ increases with \teff, and it has a tight correlation with \logg, being larger in giants than in dwarf stars
\citep{2001gray,2008takeda}.

In the series of models used by \citet{lind2008} to analyse NGC\,6397 spectra, $\xi$ decreases from $1.86 \pm 0.13$ \kms at 6250 K (TO) to $1.47 \pm 0.09$ \kms at 5450 K (SG).
In the APOGEE ASPCAP analysis \citep{garciaperez2016}, $\xi$ was fixed to a value bound to \logg through the relation $\xi = 2.478 - 0.325 \logg$. 
Along the evolutionary sequence of the cluster, this corresponds to 1.1 \kms at $\teff = 6250$\,K and 1.5 \kms at $\teff = 5250$\,K.
In GSL, $\xi$ was derived consistently from the macro-turbulence velocity and was computed for each model. According to the indications in the header of the individual files, it varies from  $\xi = 1.2$\,\kms  near the TO (near 6250\,K, at \logg=4.0\,dex) and $1.35$\,\kms at the SG (near 5450\,K, at \logg=3.0\,dex), which is almost flat.

In \AP, $\xi$ is a free parameter. To assess its importance, we compared the results obtained with the grids produced with either $\xi_1 = 1$ and  $\xi_2 = 2$ \kms.  
The mean \teff for the whole sample is not affected by the value of $\xi$: $\teff(\xi_2) - \teff(\xi_1) = 6$\,K (with a dispersion of 6\,K), but as expected, \fehe depends on $\xi$:  $\mfeh(\xi_2) - \mfeh(\xi_1) = -0.062$\,dex. 
Finally, the choice of the variation of $\xi$ along the sequence may affect the trend by $\sim 0.03$\,dex, that is, $\sim 15$\% of the observed trend. In other words, this is a noticeable effect, but it is unlikely to account for the currently observed trend.

\begin{figure*}[!ht]
     \centering
      \subfloat{\includegraphics[scale=1.00]{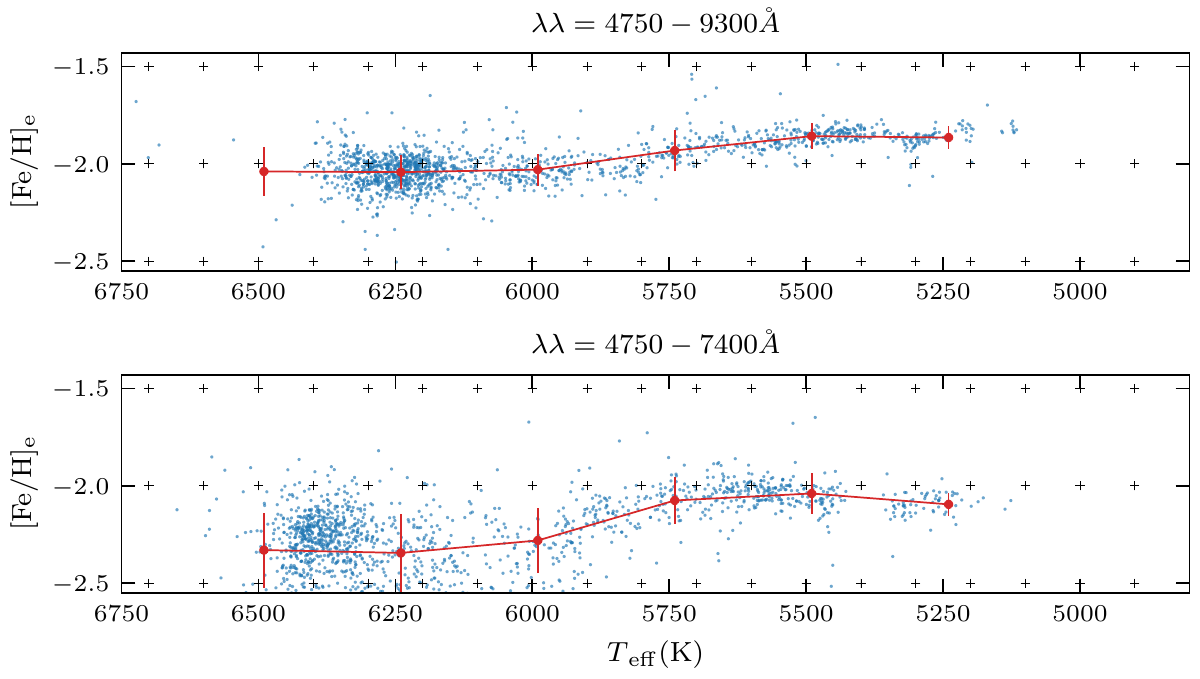}}
      \caption{\fehe vs. \teff relation obtained with GSL using different wavelength ranges of the observations. The top panel shows the relation for the full range, and the bottom panel shows the relation for the trimmed wavelength range (same as the MILES range in \RJ). The symbols and decorations are the same as in Fig~\ref{fig:relation}.}
        \label{fig:wavelengthrange}
\end{figure*}
\begin{figure*}[!ht]
     \centering
      \subfloat{\includegraphics[scale=1.00]{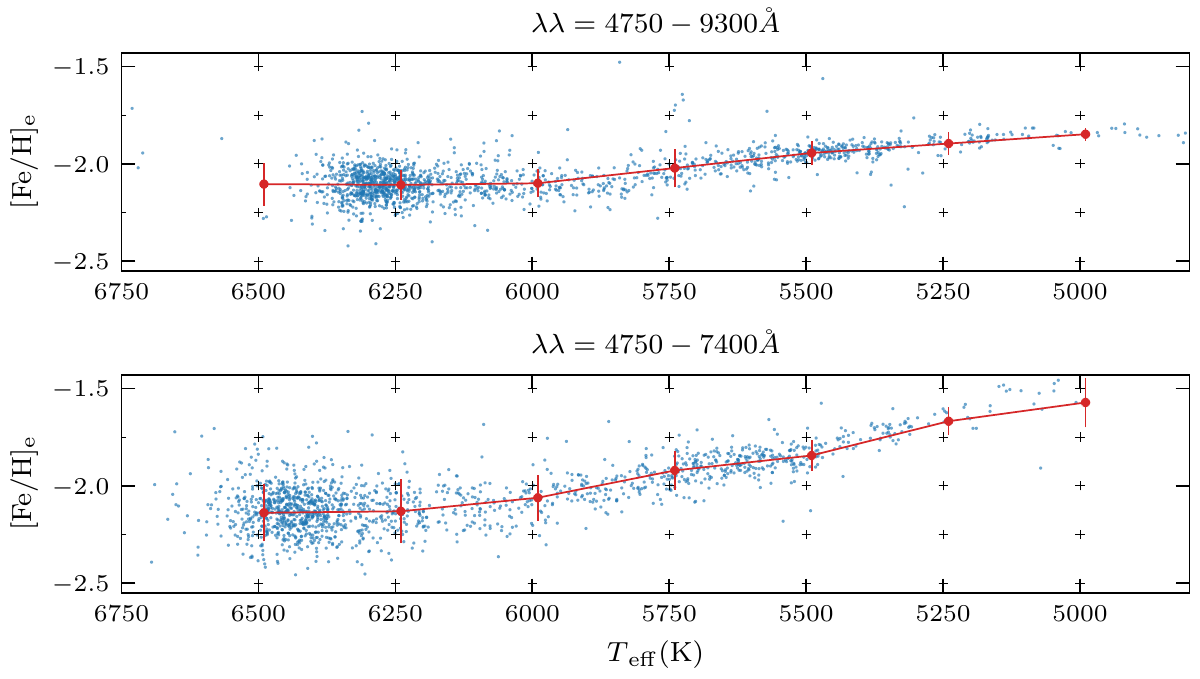}}
      \caption{\fehe vs. \teff relation obtained with \AP grids (\afe=+0.5\,dex) using different wavelength ranges of the observations: the full range (top panel), and the trimmed wavelength range (bottom panel). The symbols and decorations are the same as in Fig~\ref{fig:relation}.}
        \label{fig:wavelengthrange_AP}
\end{figure*}

\section{Discussion}
\label{sec:discussion}

We analysed the set of NGC\,6397 spectra using two synthetic libraries, and in the previous sections, we assessed how several details of the analysis may affect the precision and accuracy of the measurements. \notep{Table\,\ref{tab:summary} summarises the factors that we investigated.}
The first aspect discussed below is the accuracy, as reflected by the mean metallicity computed over the whole sample. We then focus on the trend of \feh with \teff and on the \feh spread around the cluster sequence.

\subsection{Mean metallicity}
\label{sec:wavelengths}

NGC\,6397 has $\feh = -2.02$\,dex in \citet{harris2010}. 
\citet{gratton2001} obtained high-resolution (R $\sim 43\,000$) spectra of 5 TO stars and 3 subgiants, analysed them with LTE models, and measured individual \notep{abundances} in the range $-2.10 < \feh < -2.00$\,dex, with a precision of about 0.04\,dex on each measurement.
\citet[revising the earlier analysis by \citealt{korn2007}]{nordlander2012} measured $\feh = -2.23$\,dex at the TO and $\feh = -2.13$\,dex for the SG, using 12 TO and SG stars observed with FLAMES-UVES (R=47\,000).
In a companion paper by \citet{lind2008} that used $R = 27\,000$ spectra of 116 stars, the same trend was obtained, but with an \notep{abundance} lower by $\sim 0.05$\,dex. 
\citet{lovisi2012} re-analysed the latter data and obtained $\feh = -2.12$\, dex. 
\citet{meszaros2020} used 141 giants from the APOGEE survey at R$ = 22\,500$ and obtained $\feh = -1.89$\,dex, but their metallicity scale has been documented to be $0.06 - 0.15$\,dex larger than others. This is illustrative of the uncertainty on the \feh scale due to a number of factors.
The \notep{iron abundance} at the TO may be depleted because of atomic diffusion, and the value for the SG should be more representative of the composition of the gas when the cluster formed:
we can therefore retain that the initial \notep{abundance} reported in the literature is $\feh \sim -2.1$ dex. 

This value compares well with the mean equivalent iron abundances measured with the $\alpha$-enhanced models in the full wavelength range: $-1.945$ and $-2.014$ dex, with GSL and \AP, respectively. Our values may be marginally higher, but they are still in the range of values found in earlier studies.
We then compared our results with those of \RJ by repeating the fits using the wavelength ranges explored in their paper. The two empirical libraries used by \RJ, ELODIE and MILES, are limited redwards to $\lambda = 5800$\,\AA{} and $7400$\,\AA, respectively. As these empirical libraries are made of stars from the solar neighbourhood and therefore are enhanced in $\alpha$-elements in the low-metallicity regime (similar to NGC\,6397), we consistently compared the \RJ results with those we obtained with the enhanced models.
The results for the MILES range are reported in Table~\ref{tab:feh_spread}: the mean equivalent \notep{iron abundances} with GSL and \AP are $-2.17$ and $-1.93$\,dex, respectively, compared with $-2.09$\,dex in \RJ. 
Such $\sim 0.15$\,dex differences are often observed between different series of measurements. In our case, the analysed spectra are the same and only some details in the analysis differ. We have shown in previous sections that these details are unlikely to affect the results, and the origin of the difference is certainly the choice of the set of reference spectra. For instance, the mismatch of \afe ($\afe = 0.4$ dex in GSL and 0.5 dex in \AP) can account for a $0.07$ dex difference on the \notep{equivalent iron abundance} (Sect.~\ref{sec:compa_gsl}). It should be noted that the mean solution obtained with the empirical library falls between the solutions obtained with the two synthetic libraries. We are therefore satisfied by the consistency between these results. The precision of the metallicity measurements is lower with the synthetic libraries than in \RJ ($\epsilon(\feh)$ is larger).

It should also be noted that with GSL, \mfeh decreases by $\sim 0.25$ dex when the wavelength is trimmed from the full range $\lambda\lambda = 4750 - 9300$ \AA{} to the optical domain (Fig.~\ref{fig:wavelengthrange}). The systematic is less clear with \AP (Fig.~\ref{fig:wavelengthrange_AP}): The bias on \mfeh is smaller, but it masks a decrease in \notep{abundance} for the warm stars and an increase for the cool stars. 
This difference between the different regions of the spectrum reveals the limited quality of the synthetic spectra. \notep{This includes the limitation on the atmosphere models, spectral synthesis, and atomic and molecular data, and the possible unsuitability of the abundance pattern (flat $\alpha$-enhancement).}
  \notep{Access to the bluer region of the spectrum, where  $\alpha$-enhancements opacities become more important, would certainly help to clarify this difference (this is unfortunately not possible with MUSE).}

We have discussed the degeneracy between \fehe and \afe in the full wavelength range in Sect. 5.1. Comparing the solutions reported in Table~\ref{tab:feh_spread} for the solar-scaled and enhanced models, we note that this degeneracy is a factor 2 smaller in the MILES range and 2.5 times smaller in the ELODIE range. This certainly reflects the dominance of the iron lines in the optical range.

\begin{table*} 
{\color{black} 
\caption{\notep{Summary of the different tests conducted in this work.}}
\centering

\begin{tabular}{p{0.3\textwidth} p{0.3\textwidth} p{0.3\textwidth} }
  
\toprule
\multicolumn{1}{c}{Factor} & \multicolumn{1}{c}{Test} & \multicolumn{1}{c}{Result} \\

\midrule
LSF broadening accuracy\hfill\break (Sect.~\ref{sec:lsf_effect})&
Experiment with perturbed broadening within the LSF uncertainty&
Insignificant effect\\
\\
Wavelength calibration accuracy (or line-list completeness) (Sect.~\ref{sec:lsf_effect})&
Experiment with shifted wavelengths &
Minor effect on $\tau(\feh)$\\
\\
Interpolation schema (Sect.~\ref{sec:anal_gsl_interpolation})&
Compare analyses with linear, quadratic, and cubic interpolations &
Results with cubic interpolation are marginally more precise. Artefacts with linear interpolation.\\
\\
Grid choice (Sect.~\ref{sec:compa_gsl})&
Compare analyses with the GSL and \AP grids &
Results are consistent, but the precision with \AP is marginally better\\
\\
\afe of the reference spectra (Sect.~\ref{sec:compa_gsl})&
Use reference grids with different \afe&
Significant degeneracy between \fehe and \afe\\
\\
Mesh size of the grid of reference spectra (Sect.~\ref{sec:anal_ap_coarsegrid})&
Compare solutions with full and half density grids extracted from \AP&
Minor effects with linear interpolation, insignificant effects otherwise\\
\\
Micro-turbulent velocity parameter, $\xi$ (Sect.~\ref{sec: vturb_effects})&
Compare solutions with \AP grids with $\xi=1$ and $2$\,\kms &
Minor effect on $\tau(\feh)$\\
\\
Wavelength range (Sect.~\ref{sec:wavelengths})&
Compare analyses with the full MUSE range, and with the optical region only&
Significant degradation of the precision when the range is trimmed; bias on \mfeh; different behaviour with the two grids \\
\bottomrule
\end{tabular}
\label{tab:summary}
}
\end{table*}

\subsection{Metallicity trend}

A trend of the surface metallicity along the sequence of the cluster may be due to atomic diffusion.
This is a balance between the gravitational settling and the radiative force that is counter-acted by the convective mixing. The resulting depletion of metals is larger at the TO, where the stars have thin convective layers. After the TO, these layers start to expand and to deepen, so that convection acts against atomic diffusion, re-homogenizing the matter \citep{2017salaris}. Consequently, the giant stars show the initial chemical composition at their surface. Theoretical models \citep[e.g.][]{2017dotter} predict that the Fe abundance will be $0.1-0.2$\,dex lower at the TO. 
These models predict depletions that vary for the different atomic species.

In \citet{korn2007} and \citet{nordlander2012}, a 0.1\,dex depletion of Fe at the TO compared to the SGB of NGC\,6397 was detected. They also measured similar trends for other elements (Mg, Ca, Ti, and Cr). The follow-up paper by \citet{lind2008} confirmed the higher-resolution results. It should be noted that while the former results were derived from a non-LTE analysis, the latter are derived in LTE approximation: however, given the limited \teff range of the sample, the LTE correction is not expected to induce a metallicity trend.
These trends, compared to models of atomic diffusion by \citet{2005richard}, allowed the authors to constrain the mixing parameters, which in turn could be used to derive the initial or primordial abundance of Li.
\citet{lovisi2012} independently re-analysed the latter dataset, and their results did not show the above \feh trend. Studies by other groups, albeit based on only a few stars, did show no or even opposite trends \citep{gratton2001,koch2011}.
Metallicity trends were subsequently detected in several other GCs by the same group.
In NGC\,6752, using similar data, \citet{gruyters2013,2014gruyters} found a Fe depletion of 0.08$\pm$0.06\,dex that is significant at the 1-2\,$\sigma$ level.
Finally, a similar depletion was found in M\,30 (also known as NGC\,7099) \citep{2016gruyters,gavel2021}, a more distant cluster for which measuring abundances of TO stars is challenging.

In other clusters with similar spectroscopic measurements extending to the TO, no \feh trends were observed. This is the case in M\,4 (also known as NGC\,6121, with $\feh = -1.1$\,dex) \citep{mucciarelli2011}. 

In our analysis, the trend ($\tau(\feh)$) is reported in Table~\ref{tab:feh_spread}. With the two considered synthetic grids, \fehe increases from the TO along the SG. When the analysis is carried out over the full wavelength range and models are used that are enhanced in $\alpha$-elements, we find $\tau(\feh) \sim 0.2$\,dex, which agrees with earlier results by \TOH. At first sight, it corroborates the earlier analyses of high-resolution spectra reported above. However, while we tend to have more confidence in the results based on the full spectral range because they have the best precision, we do not ignore the solutions obtained on the shortened optical range, which are notably different. They reach  $\tau(\feh) \sim 0.3 - 0.4$\,dex (Fig.~\ref{fig:wavelengthrange}~and~\ref{fig:wavelengthrange_AP}). As noted when we discussed the mean equivalent \notep{Fe abundance}, there appears to exist a difference between the optical and NIR regions, and it is certainly due to imperfections in the synthetic spectra.

The discrepancy pointed out in \RJ between the nearly constant  \notep{Fe abundance} found with empirical libraries and the rising \notep{abundance} obtained with synthetic libraries is even starker when the comparison is made over the same wavelength range. Our new analyses produced unrealistic values in some cases, in particular for the coolest stars analysed with \AP ($\fehe \sim -1.6$\,dex).

The libraries used in \RJ are sets of observations of real stars, \notep{and they provide a better spectral match with other observations than synthetic spectra do. However, these libraries had to be calibrated in \teff, \logg, and \feh using classical methods based on high spectral resolution spectroscopy, and they therefore also rely on synthetic spectra. This means that the advantage of a better spectral match may to some extend be dampened by the uncertainties linked to this indirect calibration, and in particular its inhomogeneity resulting from the many sources of measurements used to build a library. A more arguable aspect of empirical libraries, however, is their limited coverage of the parameter space.} \RJ questioned the small number of low-metallicity stars in libraries as a possible source of the discrepancy.

On the other hand, synthetic libraries \notep{entail} shortcomings in the lists of atomic or molecular lines or in the implementation of the physics in the models that are probably the origin of the poorer spectral match. They\notep{ may also cause systematics on the derived parameters that might also explain the discrepancy.}

The consistency between the solutions obtained with the two synthetic grids is an interesting point, but it does not disqualify the results of \RJ.
The two types of reference libraries have their respective strengths and weaknesses, and at this point, we cannot conclude about the accuracy of the empirical or synthetic solutions. It would be premature to interpret our full-spectrum fitting analyses of low-resolution spectra in physical terms and claim the detection of signatures of atomic diffusion.

\subsection{Cosmic metallicity spread}\label{uncertainties}

Some massive clusters, such as $\omega$ Cen, and Terzan~5, exhibit a large \feh spread \citep[see e.g.][]{2003origlia,2009ferraro} and are regarded as atypical.
Most other clusters are homogeneous concerning iron:
\citet{carretta2009} assessed the \feh spread in GCs using medium-resolution (R$~=~20\,000$) spectra of 2000 stars in 19 clusters (about 100 stars in each cluster). They found a measured spread of $0.048$\,dex, dominated by the uncertainties on the measurements to a level at which it was impossible to measure reliable intrinsic values. 

\cite{meszaros2020} more recently reached the same conclusion from a homogeneous analysis of 30 clusters in the APOGEE survey.
The compilation by \citet{bailin2019} reported 55 clusters with a possibly detected  \feh spread with a median value of 0.045\,dex. However, this is comparable to the errors on individual measurements and therefore critically depends on a precise knowledge of them (in this case, the errors are essentially derived from the dispersion of the measurements obtained on individual spectral lines; by ignoring the propagation of the random errors on the atmospheric parameters, the errors may have been under-estimated and the spread over-estimated). 
In some clusters, spreads of $\sim~0.05$\,dex were claimed, but later disputed \citep[see e.g.][in NGC\,6656]{mucciarelli2015}, reflecting the caveat of precision abundances measurements.
For $\text{}\text{about}$~40 giants in NGC\,6752, \citet{yong2013} carried out a relatively detailed abundances analysis using high-resolution spectra. They achieved an internal precision of about 0.01\,dex and measured a 0.02\,dex spread for \feh. These small spreads may be more representative of the general population of Galactic cluster, but He variation could also contribute to this apparent spread. 
In contrast, spreads in lighter elements with amplitudes reaching 0.1\,dex are widely observed \citep[e.g.][]{carretta2009}, and are certainly associated with the presence of multiple stellar populations.

In our analysis, the squared difference between the measured spread and the dispersion due to the noise and data reduction, $\Delta(\feh) = \sqrt{ \sigma(\feh)^2 - \epsilon(\feh)^2}$ , reflects the other causes of uncertainty and the intrinsic equivalent \fehe spread.
Considering the fits giving the smallest $\epsilon(\feh)$ in Table~\ref{tab:feh_spread}, those over the full wavelength range with either GSL or \AP, the residual spread is $\Delta(\feh) = 0.057$\,dex. If we consider the \RJ solution, we obtain a similar estimate: $\Delta(\feh) = 0.054$\,dex (although it is based on a shorter wavelength range, this solution has a precision comparable to that of the synthetic libraries on the full wavelength range).
This is an upper limit to the intrinsic spread in equivalent \fehe, but the consistency between the two estimates suggests that it is close to the real value.

$\Delta(\feh)$ combines the spread on the iron abundance and the effect of the spread on other elements. We discussed the degeneracy between \fehe and \afe\ above. A change in $\delta\afe$ results in a $\delta\feh$ change in $\delta\feh = a \times \delta\afe$, where according to the values in Table~\ref{tab:feh_spread}, $a = 0.58$ for GSL and $a = 0.75$ for \AP.
An intrinsic \afe spread of $\sim 0.08 - 0.1$\,dex would therefore be sufficient to explain $\Delta(\feh)$. This value is possible, and the observed residual spread does not require a spread of iron abundance.

\section{Conclusion}
\label{sec:conclusion}

We re-analysed the MUSE commissioning observations of 1061 stars in NGC\,6397 previously studied by \TOH and \RJ in order to evaluate whether such a large set of low-resolution (R$\sim 2000$) spectra may allow us to probe subtle effects in the stellar atmospheres. In particular, our aim was to determine whether variations in the surface chemical composition along the sequence of the cluster may be reliably studied or if they remain masked by effects due to the analysis procedures or to the modelling of the spectra.

Specifically, we \notep{focused on the disagreement between \TOH and \RJ} on the \fehe trend: \TOH found that the \notep{equivalent iron abundance}  rises by about 0.15 dex along the SG branch, while \RJ did not find any trend.
The two studies used the same data and similar analysis methods, but the first used as reference the GSL grid of synthetic spectra and the second used empirical libraries (ELODIE and MILES). The open questions were (i) whether the details of the analysis procedure, for instance the interpolation between spectra within the grid, affect the measured metallicty trend and (ii) whether the spectral models of stellar atmospheres have the required reliability to reveal the fine physical details.

Our analysis was made with a different code, FERRE. This code allowed us to explore the importance of the interpolation schema and different grids of models, GSL and \AP, which enabled us to explore the effect of the mesh size.

We found that (i) the interpolation schema and grid sampling are not critical with the considered grids, and (ii) GSL and \AP are essentially consistent for the measured \notep{iron abundance} trend: In both cases, we find an \notep{abundance} increase along the SG branch, which also agrees with the results of \TOH; and (iii) the precision of the measurements is mostly similar with both grids, although it is marginally better with \AP.
When it is restricted to the optical wavelength range, the precision is strongly degraded and does not rival the precision reached by \RJ using empirical libraries.

Therefore, we give a robust and unambiguous answer to the first question: The interpolation is not an issue, even for interpreting \notep{abundance} trends that are one-fivth$^{}$ or one-tenth$^{}$ of the mesh size. A linear approximation of the variations in the grid is still satisfactory, although cubic interpolations are better. 

The second question is more difficult.
The consistency between the results obtained with the two grids of models is encouraging, but does not formally imply that the results are reliable. It remains risky to simply reject the results obtained by \RJ with empirical libraries. Although these latter libraries are used at the edge of their validity space and the results stand on a limited number of reference stars with well-documented parameters, the nearly constant \notep{iron abundance} found with both MILES and ELODIE is striking. Nothing in the process biases the result towards a constant \notep{abundance}, and the probability of obtaining this result by chance is certainly low. In addition, the quality of the spectral fits is considerably better with the empirical libraries, and the \notep{abundance} uncertainties and spread are smaller.

We recall that the GSL and \AP models share some physical assumptions and ingredients. In particular, the atomic and molecular line lists have similar limitations that may explain a large part of the observed spectral residuals. The difference observed between the solutions obtained in the optical and in the optical plus NIR (see Sec.\,\ref{sec:wavelengths}) 
indicates that the models are not yet accurate enough. Metallicity biases of 0.1 or 0.2 dex are possible, and the divergence of the \notep{iron abundance} trend when the wavelength range is shortened (in particular with \AP) casts some doubts on the reality of this trend.

\notep{In contrast, we found the trend to be relatively robust toward the choice of the micro-turbulence velocity parameter and of its variation along the cluster sequence. The possible effect is only $\sim$0.03\,dex on $\tau$($\feh$).}

The difference between synthetic and empirical libraries is not explained, and there are still some open questions that need to be further investigated in the future. The use of the X-Shooter Spectral Library \citep{gonneau2020,verro21} to fit the whole wavelength range may help to shed light on the current difference observed between optical and NIR.
Our dataset comes from short exposures, and therefore the achieved S/N is not the best that could be obtained with the instrument. The observed spread is definitely dominated by the effect of the noise. With better observations, the main source of error may be lowered, which would improve the determination of the intrinsic equivalent \notep{iron abundance} spread. It may also allow us to separate the \afe and \feh spreads. Improving the synthetic models is a more difficult and critical issue. One route could be the empirical calibration of the atomic and molecular line lists \citep{martins2014,shetrone2015,franchini18}, but this may mask some more fundamental questions in the stellar models, such as the effects of non-LTE or 3D treatment.

\begin{acknowledgements}
We thank the anonymous referee for very helpful comments and suggestions. We thank Tim-Oliver Husser who shared with us the tables with the results of the H16 analysis, and the MUSE team for making public the extracted spectra on their website. We are grateful to Lucimara Martins, Ariane Lan\c{c}on, Anke Arentsen, Paula Coelho, and Rashi Jain for fruitful discussions. We acknowledge the LABEX Lyon Institute of Origins (ANR-10-LABX-0066) of the Université de Lyon for its support within the program "Investissements d’Avenir" (ANR-11-IDEX-0007) of the French government operated by the National Research Agency (ANR), IUCAA and CRAL for their support for the IFAS summer school where this work was started.
  This work was supported in part by grant ANR-19-CE31-0022 (POPSYCLE) of the French ANR.
\end{acknowledgements}

%
\bibliographystyle{aa} 
\interlinepenalty=10000
\bibliography{paper.bib} 

\begin{thebibliography}{62}
\expandafter\ifx\csname natexlab\endcsname\relax\def\natexlab#1{#1}\fi

\bibitem[{{Allende-Prieto} \& {Apogee Team}(2015)}]{allende-prieto15}
{Allende-Prieto}, C. \& {Apogee Team}. 2015, in American Astronomical Society
  Meeting Abstracts, Vol. 225, American Astronomical Society Meeting Abstracts
  \#225, 422.07

\bibitem[{{Allende Prieto} {et~al.}(2018){Allende Prieto}, {Koesterke},
  {Hubeny}, {Bautista}, {Barklem}, \& {Nahar}}]{allende-prieto18}
{Allende Prieto}, C., {Koesterke}, L., {Hubeny}, I., {et~al.} 2018, \aap, 618,
  A25

\bibitem[{{Anderson} {et~al.}(2008){Anderson}, {Sarajedini}, {Bedin}, {King},
  {Piotto}, {Reid}, {Siegel}, {Majewski}, {Paust}, {Aparicio}, {Milone},
  {Chaboyer}, \& {Rosenberg}}]{anderson2008}
{Anderson}, J., {Sarajedini}, A., {Bedin}, L.~R., {et~al.} 2008, \aj, 135, 2055

\bibitem[{{Asplund} {et~al.}(2009){Asplund}, {Grevesse}, {Sauval}, \&
  {Scott}}]{2009asplund}
{Asplund}, M., {Grevesse}, N., {Sauval}, A.~J., \& {Scott}, P. 2009, \araa, 47,
  481

\bibitem[{{Auer}(2003)}]{auer2003}
{Auer}, L. 2003, in Astronomical Society of the Pacific Conference Series, Vol.
  288, Stellar Atmosphere Modeling, ed. I.~{Hubeny}, D.~{Mihalas}, \&
  K.~{Werner}, 3

\bibitem[{{Bacon} {et~al.}(2010){Bacon}, {Accardo}, {Adjali}, {Anwand},
  {Bauer}, {Biswas}, {Blaizot}, {Boudon}, {Brau-Nogue}, {Brinchmann},
  {Caillier}, {Capoani}, {Carollo}, {Contini}, {Couderc}, {Daguis{\'e}},
  {Deiries}, {Delabre}, {Dreizler}, {Dubois}, {Dupieux}, {Dupuy}, {Emsellem},
  {Fechner}, {Fleischmann}, {Fran{\c{c}}ois}, {Gallou}, {Gharsa}, {Glindemann},
  {Gojak}, {Guiderdoni}, {Hansali}, {Hahn}, {Jarno}, {Kelz}, {Koehler},
  {Kosmalski}, {Laurent}, {Le Floch}, {Lilly}, {Lizon}, {Loupias}, {Manescau},
  {Monstein}, {Nicklas}, {Olaya}, {Pares}, {Pasquini}, {P{\'e}contal-Rousset},
  {Pell{\'o}}, {Petit}, {Popow}, {Reiss}, {Remillieux}, {Renault}, {Roth},
  {Rupprecht}, {Serre}, {Schaye}, {Soucail}, {Steinmetz}, {Streicher}, {Stuik},
  {Valentin}, {Vernet}, {Weilbacher}, {Wisotzki}, \& {Yerle}}]{2010bacon}
{Bacon}, R., {Accardo}, M., {Adjali}, L., {et~al.} 2010, in Society of
  Photo-Optical Instrumentation Engineers (SPIE) Conference Series, Vol. 7735,
  Ground-based and Airborne Instrumentation for Astronomy III, ed. I.~S.
  {McLean}, S.~K. {Ramsay}, \& H.~{Takami}, 773508

\bibitem[{{Bailin}(2019)}]{bailin2019}
{Bailin}, J. 2019, \apjs, 245, 5

\bibitem[{{Bressan} {et~al.}(2012){Bressan}, {Marigo}, {Girardi}, {Salasnich},
  {Dal Cero}, {Rubele}, \& {Nanni}}]{bressan12}
{Bressan}, A., {Marigo}, P., {Girardi}, L., {et~al.} 2012, \mnras, 427, 127

\bibitem[{{Carretta} {et~al.}(2009){Carretta}, {Bragaglia}, {Gratton},
  {D'Orazi}, \& {Lucatello}}]{carretta2009}
{Carretta}, E., {Bragaglia}, A., {Gratton}, R., {D'Orazi}, V., \& {Lucatello},
  S. 2009, \aap, 508, 695

\bibitem[{{Chapman}(1917)}]{chapman1917}
{Chapman}, S. 1917, \mnras, 77, 540

\bibitem[{{Dotter} {et~al.}(2017){Dotter}, {Conroy}, {Cargile}, \&
  {Asplund}}]{2017dotter}
{Dotter}, A., {Conroy}, C., {Cargile}, P., \& {Asplund}, M. 2017, \apj, 840, 99

\bibitem[{{Ferraro} {et~al.}(2009){Ferraro}, {Dalessandro}, {Mucciarelli},
  {Beccari}, {Rich}, {Origlia}, {Lanzoni}, {Rood}, {Valenti}, {Bellazzini},
  {Ransom}, \& {Cocozza}}]{2009ferraro}
{Ferraro}, F.~R., {Dalessandro}, E., {Mucciarelli}, A., {et~al.} 2009, \nat,
  462, 483

\bibitem[{{Franchini} {et~al.}(2018){Franchini}, {Morossi}, {Di Marcantonio},
  {Chavez}, {Gilmore}, {Randich}, {Flaccomio}, {Koposov}, {Korn}, {Bayo},
  {Carraro}, {Casey}, {Franciosini}, {Hourihane}, {Jofr{\'e}}, {Lardo},
  {Lewis}, {Magrini}, {Morbidelli}, {Sacco}, {Worley}, \&
  {Zwitter}}]{franchini18}
{Franchini}, M., {Morossi}, C., {Di Marcantonio}, P., {et~al.} 2018, \apj, 862,
  146

\bibitem[{{Garc{\'\i}a P{\'e}rez} {et~al.}(2016){Garc{\'\i}a P{\'e}rez},
  {Allende Prieto}, {Holtzman}, {Shetrone}, {M{\'e}sz{\'a}ros}, {Bizyaev},
  {Carrera}, {Cunha}, {Garc{\'\i}a-Hern{\'a}ndez}, {Johnson}, {Majewski},
  {Nidever}, {Schiavon}, {Shane}, {Smith}, {Sobeck}, {Troup}, {Zamora},
  {Weinberg}, {Bovy}, {Eisenstein}, {Feuillet}, {Frinchaboy}, {Hayden},
  {Hearty}, {Nguyen}, {O'Connell}, {Pinsonneault}, {Wilson}, \&
  {Zasowski}}]{garciaperez2016}
{Garc{\'\i}a P{\'e}rez}, A.~E., {Allende Prieto}, C., {Holtzman}, J.~A.,
  {et~al.} 2016, \aj, 151, 144

\bibitem[{{Gavel} {et~al.}(2021){Gavel}, {Gruyters}, {Heiter}, {Korn},
  {Nordlander}, {Scheutwinkel}, \& {Richard}}]{gavel2021}
{Gavel}, A., {Gruyters}, P., {Heiter}, U., {et~al.} 2021, \aap, 652, A75

\bibitem[{{Gilmore} {et~al.}(2012){Gilmore}, {Randich}, {Asplund}, {Binney},
  {Bonifacio}, {Drew}, {Feltzing}, {Ferguson}, {Jeffries}, {Micela},
  {Negueruela}, {Prusti}, {Rix}, {Vallenari}, {Alfaro}, {Allende-Prieto},
  {Babusiaux}, {Bensby}, {Blomme}, {Bragaglia}, {Flaccomio}, {Fran{\c{c}}ois},
  {Irwin}, {Koposov}, {Korn}, {Lanzafame}, {Pancino}, {Paunzen},
  {Recio-Blanco}, {Sacco}, {Smiljanic}, {Van Eck}, {Walton}, {Aden}, {Aerts},
  {Affer}, {Alcala}, {Altavilla}, {Alves}, {Antoja}, {Arenou}, {Argiroffi},
  {Asensio Ramos}, {Bailer-Jones}, {Balaguer-Nunez}, {Bayo}, {Barbuy},
  {Barisevicius}, {Barrado y Navascues}, {Battistini}, {Bellas Velidis},
  {Bellazzini}, {Belokurov}, {Bergemann}, {Bertelli}, {Biazzo}, {Bienayme},
  {Bland-Hawthorn}, {Boeche}, {Bonito}, {Boudreault}, {Bouvier}, {Brandao},
  {Brown}, {de Bruijne}, {Burleigh}, {Caballero}, {Caffau}, {Calura},
  {Capuzzo-Dolcetta}, {Caramazza}, {Carraro}, {Casagrande}, {Casewell},
  {Chapman}, {Chiappini}, {Chorniy}, {Christlieb}, {Cignoni}, {Cocozza},
  {Colless}, {Collet}, {Collins}, {Correnti}, {Covino}, {Crnojevic}, {Cropper},
  {Cunha}, {Damiani}, {David}, {Delgado}, {Duffau}, {Edvardsson}, {Eldridge},
  {Enke}, {Eriksson}, {Evans}, {Eyer}, {Famaey}, {Fellhauer}, {Ferreras},
  {Figueras}, {Fiorentino}, {Flynn}, {Folha}, {Franciosini}, {Frasca},
  {Freeman}, {Fremat}, {Friel}, {Gaensicke}, {Gameiro}, {Garzon}, {Geier},
  {Geisler}, {Gerhard}, {Gibson}, {Gomboc}, {Gomez}, {Gonzalez-Fernandez},
  {Gonzalez Hernandez}, {Gosset}, {Grebel}, {Greimel}, {Groenewegen},
  {Grundahl}, {Guarcello}, {Gustafsson}, {Hadrava}, {Hatzidimitriou}, {Hambly},
  {Hammersley}, {Hansen}, {Haywood}, {Heber}, {Heiter}, {Held}, {Helmi},
  {Hensler}, {Herrero}, {Hill}, {Hodgkin}, {Huelamo}, {Huxor}, {Ibata},
  {Jackson}, {de Jong}, {Jonker}, {Jordan}, {Jordi}, {Jorissen}, {Katz},
  {Kawata}, {Keller}, {Kharchenko}, {Klement}, {Klutsch}, {Knude}, {Koch},
  {Kochukhov}, {Kontizas}, {Koubsky}, {Lallement}, {de Laverny}, {van Leeuwen},
  {Lemasle}, {Lewis}, {Lind}, {Lindstrom}, {Lobel}, {Lopez Santiago}, {Lucas},
  {Ludwig}, {Lueftinger}, {Magrini}, {Maiz Apellaniz}, {Maldonado}, {Marconi},
  {Marino}, {Martayan}, {Martinez-Valpuesta}, {Matijevic}, {McMahon},
  {Messina}, {Meyer}, {Miglio}, {Mikolaitis}, {Minchev}, {Minniti}, {Moitinho},
  {Momany}, {Monaco}, {Montalto}, {Monteiro}, {Monier}, {Montes}, {Mora},
  {Moraux}, {Morel}, {Mowlavi}, {Mucciarelli}, {Munari}, {Napiwotzki},
  {Nardetto}, {Naylor}, {Naze}, {Nelemans}, {Okamoto}, {Ortolani}, {Pace},
  {Palla}, {Palous}, {Parker}, {Penarrubia}, {Pillitteri}, {Piotto}, {Posbic},
  {Prisinzano}, {Puzeras}, {Quirrenbach}, {Ragaini}, {Read}, {Read}, {Reyle},
  {De Ridder}, {Robichon}, {Robin}, {Roeser}, {Romano}, {Royer}, {Ruchti},
  {Ruzicka}, {Ryan}, {Ryde}, {Santos}, {Sanz Forcada}, {Sarro Baro},
  {Sbordone}, {Schilbach}, {Schmeja}, {Schnurr}, {Schoenrich}, {Scholz},
  {Seabroke}, {Sharma}, {De Silva}, {Smith}, {Solano}, {Sordo}, {Soubiran},
  {Sousa}, {Spagna}, {Steffen}, {Steinmetz}, {Stelzer}, {Stempels},
  {Tabernero}, {Tautvaisiene}, {Thevenin}, {Torra}, {Tosi}, {Tolstoy}, {Turon},
  {Walker}, {Wambsganss}, {Worley}, {Venn}, {Vink}, {Wyse}, {Zaggia},
  {Zeilinger}, {Zoccali}, {Zorec}, {Zucker}, {Zwitter}, \& {Gaia-ESO Survey
  Team}}]{2012gilmore}
{Gilmore}, G., {Randich}, S., {Asplund}, M., {et~al.} 2012, The Messenger, 147,
  25

\bibitem[{{Gonneau} {et~al.}(2020){Gonneau}, {Lyubenova}, {Lan{\c{c}}on},
  {Trager}, {Peletier}, {Arentsen}, {Chen}, {Coelho}, {Dries},
  {Falc{\'o}n-Barroso}, {Prugniel}, {S{\'a}nchez-Bl{\'a}zquez}, {Vazdekis}, \&
  {Verro}}]{gonneau2020}
{Gonneau}, A., {Lyubenova}, M., {Lan{\c{c}}on}, A., {et~al.} 2020, \aap, 634,
  A133

\bibitem[{{Gratton} {et~al.}(2001){Gratton}, {Bonifacio}, {Bragaglia},
  {Carretta}, {Castellani}, {Centurion}, {Chieffi}, {Claudi}, {Clementini},
  {D'Antona}, {Desidera}, {Fran{\c c}ois}, {Grundahl}, {Lucatello}, {Molaro},
  {Pasquini}, {Sneden}, {Spite}, \& {Straniero}}]{gratton2001}
{Gratton}, R.~G., {Bonifacio}, P., {Bragaglia}, A., {et~al.} 2001, \aap, 369,
  87

\bibitem[{{Gratton} {et~al.}(2012){Gratton}, {Carretta}, \&
  {Bragaglia}}]{gratton12}
{Gratton}, R.~G., {Carretta}, E., \& {Bragaglia}, A. 2012, \aapr, 20, 50

\bibitem[{{Gray} {et~al.}(2001){Gray}, {Graham}, \& {Hoyt}}]{2001gray}
{Gray}, R.~O., {Graham}, P.~W., \& {Hoyt}, S.~R. 2001, \aj, 121, 2159

\bibitem[{{Gruyters} {et~al.}(2013){Gruyters}, {Korn}, {Richard}, {Grundahl},
  {Collet}, {Mashonkina}, {Osorio}, \& {Barklem}}]{gruyters2013}
{Gruyters}, P., {Korn}, A.~J., {Richard}, O., {et~al.} 2013, \aap, 555, A31

\bibitem[{{Gruyters} {et~al.}(2016){Gruyters}, {Lind}, {Richard}, {Grundahl},
  {Asplund}, {Casagrande}, {Charbonnel}, {Milone}, {Primas}, \&
  {Korn}}]{2016gruyters}
{Gruyters}, P., {Lind}, K., {Richard}, O., {et~al.} 2016, \aap, 589, A61

\bibitem[{{Gruyters} {et~al.}(2014){Gruyters}, {Nordlander}, \&
  {Korn}}]{2014gruyters}
{Gruyters}, P., {Nordlander}, T., \& {Korn}, A.~J. 2014, \aap, 567, A72

\bibitem[{{Gustafsson} {et~al.}(2008){Gustafsson}, {Edvardsson}, {Eriksson},
  {J{\o}rgensen}, {Nordlund}, \& {Plez}}]{2008gustafsson}
{Gustafsson}, B., {Edvardsson}, B., {Eriksson}, K., {et~al.} 2008, \aap, 486,
  951

\bibitem[{{Harris}(2010)}]{harris2010}
{Harris}, W.~E. 2010, arXiv e-prints, arXiv:1012.3224

\bibitem[{{Hauschildt} \& {Baron}(1999)}]{hauschildt99}
{Hauschildt}, P.~H. \& {Baron}, E. 1999, Journal of Computational and Applied
  Mathematics, 109, 41

\bibitem[{{Husser} {et~al.}(2016){Husser}, {Kamann}, {Dreizler}, {Wendt},
  {Wulff}, {Bacon}, {Wisotzki}, {Brinchmann}, {Weilbacher}, {Roth}, \&
  {Monreal-Ibero}}]{husser16}
{Husser}, T.-O., {Kamann}, S., {Dreizler}, S., {et~al.} 2016, \aap, 588, A148

\bibitem[{{Husser} {et~al.}(2013){Husser}, {Wende-von Berg}, {Dreizler},
  {Homeier}, {Reiners}, {Barman}, \& {Hauschildt}}]{2013Husser}
{Husser}, T.~O., {Wende-von Berg}, S., {Dreizler}, S., {et~al.} 2013, \aap,
  553, A6

\bibitem[{{Jain} {et~al.}(2020){Jain}, {Prugniel}, {Martins}, \&
  {Lan{\c{c}}on}}]{jain20}
{Jain}, R., {Prugniel}, P., {Martins}, L., \& {Lan{\c{c}}on}, A. 2020, \aap,
  635, A161

\bibitem[{{Kamann} {et~al.}(2013){Kamann}, {Wisotzki}, \& {Roth}}]{kamann13}
{Kamann}, S., {Wisotzki}, L., \& {Roth}, M.~M. 2013, \aap, 549, A71

\bibitem[{{Koch} \& {McWilliam}(2011)}]{koch2011}
{Koch}, A. \& {McWilliam}, A. 2011, \aj, 142, 63

\bibitem[{{Koesterke}(2009)}]{koesterke2009}
{Koesterke}, L. 2009, in American Institute of Physics Conference Series, Vol.
  1171, Recent Directions in Astrophysical Quantitative Spectroscopy and
  Radiation Hydrodynamics, ed. I.~{Hubeny}, J.~M. {Stone}, K.~{MacGregor}, \&
  K.~{Werner}, 73--84

\bibitem[{{Koleva} {et~al.}(2009){Koleva}, {Prugniel}, {Bouchard}, \&
  {Wu}}]{koleva09}
{Koleva}, M., {Prugniel}, P., {Bouchard}, A., \& {Wu}, Y. 2009, \aap, 501, 1269

\bibitem[{{Korn} {et~al.}(2007){Korn}, {Grundahl}, {Richard}, {Mashonkina},
  {Barklem}, {Collet}, {Gustafsson}, \& {Piskunov}}]{korn2007}
{Korn}, A.~J., {Grundahl}, F., {Richard}, O., {et~al.} 2007, \apj, 671, 402

\bibitem[{{Kurucz}(2009)}]{2009kurucz}
{Kurucz}, R.~L. 2009, in American Institute of Physics Conference Series, Vol.
  1171, Recent Directions in Astrophysical Quantitative Spectroscopy and
  Radiation Hydrodynamics, ed. I.~{Hubeny}, J.~M. {Stone}, K.~{MacGregor}, \&
  K.~{Werner}, 43--51

\bibitem[{{Lan{\c{c}}on} {et~al.}(2021){Lan{\c{c}}on}, {Gonneau}, {Verro},
  {Prugniel}, {Arentsen}, {Trager}, {Peletier}, {Chen}, {Coelho},
  {Falc{\'o}n-Barroso}, {Hauschildt}, {Husser}, {Jain}, {Lyubenova}, {Martins},
  {S{\'a}nchez Bl{\'a}zquez}, \& {Vazdekis}}]{2021lancon}
{Lan{\c{c}}on}, A., {Gonneau}, A., {Verro}, K., {et~al.} 2021, \aap, 649, A97

\bibitem[{{Lee} {et~al.}(2008){Lee}, {Beers}, {Sivarani}, {Allende Prieto},
  {Koesterke}, {Wilhelm}, {Re Fiorentin}, {Bailer-Jones}, {Norris}, {Rockosi},
  {Yanny}, {Newberg}, {Covey}, {Zhang}, \& {Luo}}]{lee2008a}
{Lee}, Y.~S., {Beers}, T.~C., {Sivarani}, T., {et~al.} 2008, \aj, 136, 2022

\bibitem[{{Leibniz}(1765)}]{leibniz}
{Leibniz}, G.~W. 1765, New Essay on Human Understanding (Book IV, Chapter 16)

\bibitem[{{Lind} {et~al.}(2008){Lind}, {Korn}, {Barklem}, \&
  {Grundahl}}]{lind2008}
{Lind}, K., {Korn}, A.~J., {Barklem}, P.~S., \& {Grundahl}, F. 2008, \aap, 490,
  777

\bibitem[{{Lovisi} {et~al.}(2012){Lovisi}, {Mucciarelli}, {Lanzoni}, {Ferraro},
  {Gratton}, {Dalessandro}, \& {Contreras Ramos}}]{lovisi2012}
{Lovisi}, L., {Mucciarelli}, A., {Lanzoni}, B., {et~al.} 2012, \apj, 754, 91

\bibitem[{{Martins} \& {Coelho}(2007)}]{2007martins}
{Martins}, L.~P. \& {Coelho}, P. 2007, \mnras, 381, 1329

\bibitem[{{Martins} {et~al.}(2014){Martins}, {Coelho}, {Caproni}, \&
  {Vitoriano}}]{martins2014}
{Martins}, L.~P., {Coelho}, P., {Caproni}, A., \& {Vitoriano}, R. 2014, \mnras,
  442, 1294

\bibitem[{{Martins} {et~al.}(2019){Martins}, {Lima-Dias}, {Coelho}, \&
  {Lagan{\'a}}}]{martins19}
{Martins}, L.~P., {Lima-Dias}, C., {Coelho}, P. R.~T., \& {Lagan{\'a}}, T.~F.
  2019, \mnras, 484, 2388

\bibitem[{{M{\'e}sz{\'a}ros} \& {Allende Prieto}(2013)}]{meszaros13b}
{M{\'e}sz{\'a}ros}, S. \& {Allende Prieto}, C. 2013, \mnras, 430, 3285

\bibitem[{{M{\'e}sz{\'a}ros} {et~al.}(2012){M{\'e}sz{\'a}ros}, {Allende
  Prieto}, {Edvardsson}, {Castelli}, {Garc{\'\i}a P{\'e}rez}, {Gustafsson},
  {Majewski}, {Plez}, {Schiavon}, {Shetrone}, \& {de Vicente}}]{meszaros12}
{M{\'e}sz{\'a}ros}, S., {Allende Prieto}, C., {Edvardsson}, B., {et~al.} 2012,
  \aj, 144, 120

\bibitem[{{M{\'e}sz{\'a}ros} {et~al.}(2015){M{\'e}sz{\'a}ros}, {Martell},
  {Shetrone}, {Lucatello}, {Troup}, {Bovy}, {Cunha},
  {Garc{\'\i}a-Hern{\'a}ndez}, {Overbeek}, {Allende Prieto}, {Beers},
  {Frinchaboy}, {Garc{\'\i}a P{\'e}rez}, {Hearty}, {Holtzman}, {Majewski},
  {Nidever}, {Schiavon}, {Schneider}, {Sobeck}, {Smith}, {Zamora}, \&
  {Zasowski}}]{meszaros15}
{M{\'e}sz{\'a}ros}, S., {Martell}, S.~L., {Shetrone}, M., {et~al.} 2015, \aj,
  149, 153

\bibitem[{{M{\'e}sz{\'a}ros} {et~al.}(2020){M{\'e}sz{\'a}ros}, {Masseron},
  {Garc{\'\i}a-Hern{\'a}ndez}, {Allende Prieto}, {Beers}, {Bizyaev},
  {Chojnowski}, {Cohen}, {Cunha}, {Dell'Agli}, {Ebelke},
  {Fern{\'a}ndez-Trincado}, {Frinchaboy}, {Geisler}, {Hasselquist}, {Hearty},
  {Holtzman}, {Johnson}, {Lane}, {Lacerna}, {Longa-Pe{\~n}a}, {Majewski},
  {Martell}, {Minniti}, {Nataf}, {Nidever}, {Pan}, {Schiavon}, {Shetrone},
  {Smith}, {Sobeck}, {Stringfellow}, {Szigeti}, {Tang}, {Wilson}, \&
  {Zamora}}]{meszaros2020}
{M{\'e}sz{\'a}ros}, S., {Masseron}, T., {Garc{\'\i}a-Hern{\'a}ndez}, D.~A.,
  {et~al.} 2020, \mnras, 492, 1641

\bibitem[{{Mucciarelli}(2011)}]{2011mucciarelli}
{Mucciarelli}, A. 2011, \aap, 528, A44

\bibitem[{{Mucciarelli} {et~al.}(2015){Mucciarelli}, {Lapenna}, {Massari},
  {Pancino}, {Stetson}, {Ferraro}, {Lanzoni}, \& {Lardo}}]{mucciarelli2015}
{Mucciarelli}, A., {Lapenna}, E., {Massari}, D., {et~al.} 2015, \apj, 809, 128

\bibitem[{{Mucciarelli} {et~al.}(2011){Mucciarelli}, {Salaris}, {Lovisi},
  {Ferraro}, {Lanzoni}, {Lucatello}, \& {Gratton}}]{mucciarelli2011}
{Mucciarelli}, A., {Salaris}, M., {Lovisi}, L., {et~al.} 2011, \mnras, 412, 81

\bibitem[{{Nordlander} {et~al.}(2012){Nordlander}, {Korn}, {Richard}, \&
  {Lind}}]{nordlander2012}
{Nordlander}, T., {Korn}, A.~J., {Richard}, O., \& {Lind}, K. 2012, \apj, 753,
  48

\bibitem[{{Origlia} {et~al.}(2003){Origlia}, {Ferraro}, {Bellazzini}, \&
  {Pancino}}]{2003origlia}
{Origlia}, L., {Ferraro}, F.~R., {Bellazzini}, M., \& {Pancino}, E. 2003, \apj,
  591, 916

\bibitem[{Press {et~al.}(1992)Press, Teukolsky, Vetterling, \&
  Flannery}]{press92}
Press, W.~H., Teukolsky, S.~A., Vetterling, W.~T., \& Flannery, B.~P. 1992,
  Numerical Recipes in Fortran 77, 2nd edn. (Cambridge, USA: Cambridge
  University Press)

\bibitem[{{Prugniel} \& {Soubiran}(2001)}]{2001prugniel}
{Prugniel}, P. \& {Soubiran}, C. 2001, \aap, 369, 1048

\bibitem[{{Richard} {et~al.}(2005){Richard}, {Michaud}, \&
  {Richer}}]{2005richard}
{Richard}, O., {Michaud}, G., \& {Richer}, J. 2005, \apj, 619, 538

\bibitem[{{Salaris} \& {Cassisi}(2017)}]{2017salaris}
{Salaris}, M. \& {Cassisi}, S. 2017, Royal Society Open Science, 4, 170192

\bibitem[{{S{\'a}nchez-Bl{\'a}zquez} {et~al.}(2006){S{\'a}nchez-Bl{\'a}zquez},
  {Peletier}, {Jim{\'e}nez-Vicente}, {Cardiel}, {Cenarro},
  {Falc{\'o}n-Barroso}, {Gorgas}, {Selam}, \&
  {Vazdekis}}]{sanchez-blazquez2006}
{S{\'a}nchez-Bl{\'a}zquez}, P., {Peletier}, R.~F., {Jim{\'e}nez-Vicente}, J.,
  {et~al.} 2006, \mnras, 371, 703

\bibitem[{{Shetrone} {et~al.}(2015){Shetrone}, {Bizyaev}, {Lawler}, {Allende
  Prieto}, {Johnson}, {Smith}, {Cunha}, {Holtzman}, {Garc{\'\i}a P{\'e}rez},
  {M{\'e}sz{\'a}ros}, {Sobeck}, {Zamora}, {Garc{\'\i}a-Hern{\'a}ndez}, {Souto},
  {Chojnowski}, {Koesterke}, {Majewski}, \& {Zasowski}}]{shetrone2015}
{Shetrone}, M., {Bizyaev}, D., {Lawler}, J.~E., {et~al.} 2015, \apjs, 221, 24

\bibitem[{{Takeda} {et~al.}(2008){Takeda}, {Sato}, \& {Murata}}]{2008takeda}
{Takeda}, Y., {Sato}, B., \& {Murata}, D. 2008, \pasj, 60, 781

\bibitem[{{Verro} {et~al.}(2021){Verro}, {Trager}, {Peletier}, {Lan{\c{c}}on},
  {Gonneau}, {Vazdekis}, {Prugniel}, {Chen}, {Coelho},
  {S{\'a}nchez-Bl{\'a}zquez}, {Martins}, {Arentsen}, {Lyubenova},
  {Falc{\'o}n-Barroso}, \& {Dries}}]{verro21}
{Verro}, K., {Trager}, S.~C., {Peletier}, R.~F., {et~al.} 2021, arXiv e-prints,
  arXiv:2110.10188

\bibitem[{{Wu} {et~al.}(2011){Wu}, {Singh}, {Prugniel}, {Gupta}, \&
  {Koleva}}]{wu2011}
{Wu}, Y., {Singh}, H.~P., {Prugniel}, P., {Gupta}, R., \& {Koleva}, M. 2011,
  \aap, 525, A71

\bibitem[{{Yong} {et~al.}(2013){Yong}, {Mel{\'e}ndez}, {Grundahl}, {Roederer},
  {Norris}, {Milone}, {Marino}, {Coelho}, {McArthur}, {Lind}, {Collet}, \&
  {Asplund}}]{yong2013}
{Yong}, D., {Mel{\'e}ndez}, J., {Grundahl}, F., {et~al.} 2013, \mnras, 434,
  3542

\end{thebibliography}


\newpage

\begin{appendix}

\section{Discontinuity and outliers in GSL}
\label{sec:facitsaltum}

\begin{figure*}[!htb]
\centering
     \subfloat{\includegraphics[scale=1.0, trim=0 9 0 25, clip]{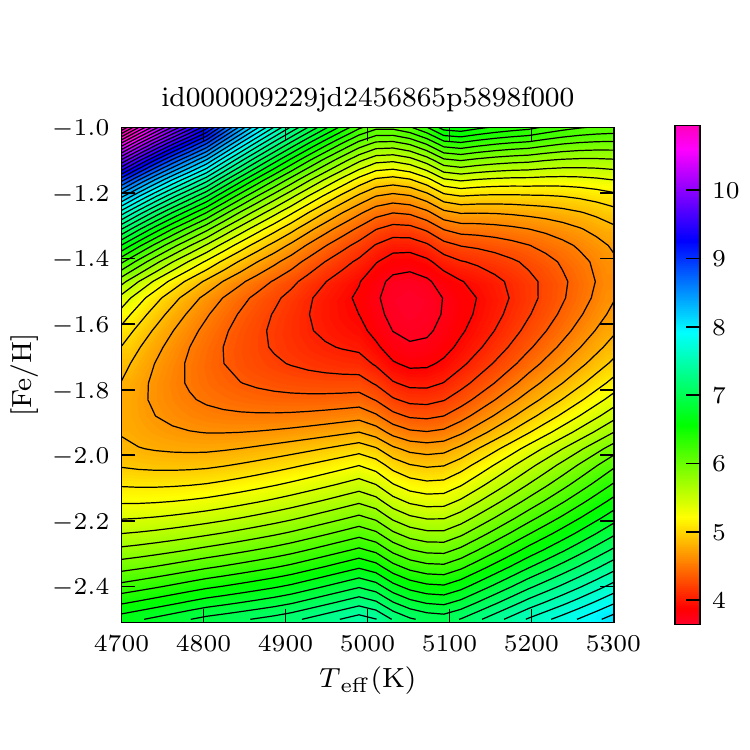}}
     \subfloat{\includegraphics[scale=1.0, trim=0 9 0 25, clip]{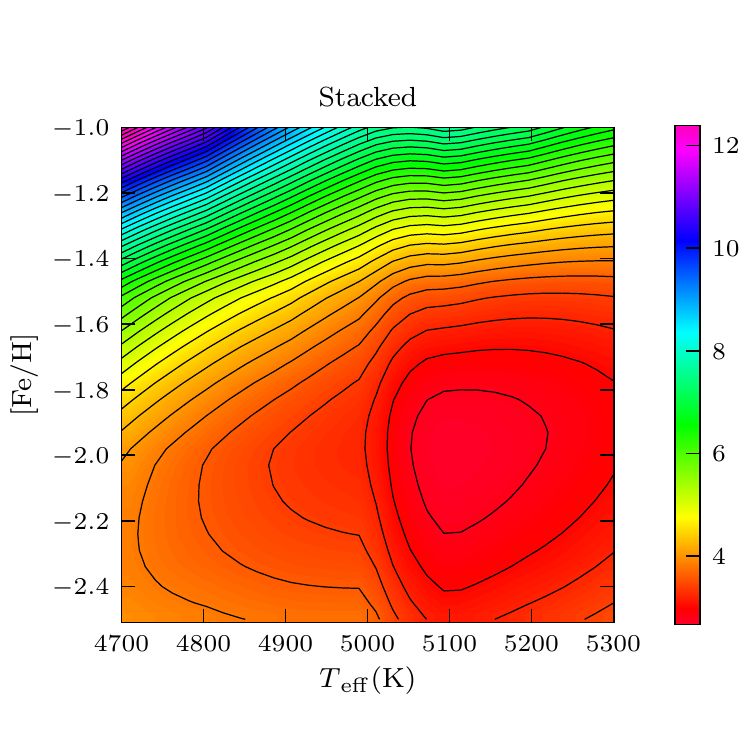}} 
     \caption{$\chi^2$ map with GSL in the region of the $\teff = 5000$\,K discontinuity. On the right, we show the spectrum ID = {\tt id000009229jd2456865p5898f000} in the full wavelength range. On the left, we show a stacked map for the 214 spectra with $4300 < \teff < 5300$\,K, selected from \RJ.
     }
     \label{fig:chi2}
\end{figure*}

Natura non facit saltum \citep[nature does not make jumps, ][]{leibniz} expresses a ``continuity principle'' that has been called along the centuries to justify or describe important theories in various fields of sciences, from the infinitesimal calculus to the evolution of biological organisms. If not a strong physical law, this is definitely an assumption behind interpolating in a grid of models to determine physical parameters.
In the considered region of parameters, the stellar spectra are continuous with respect to \teff, \logg, and chemical composition, but artificial discontinuities, irregularities, or outlier spectra may arise from the modelling.
The irregular distribution of the measurements along the cluster sequence (see \TOH their figure 8 and \RJ their figure 1) suggests such discontinuities: there are gaps and concentrations in the \teff coverage, in particular, near  $\teff = 5000$\,K. Tim-Oliver Husser reported suspicions of problems in GSL in this region for giant stars (private communication).

Therefore, we explored the grid in search for outliers or discontinuities. 
For each \logg, \feh, and \afe, we computed the difference between a spectrum on the grid and a spectrum that was linearly interpolated using the preceding and following spectra along the temperature sequence. This allowed us to clearly detect a sharp discontinuity separating the region with $\teff \leq 5000$\, K from the rest (this affects the entire range of gravities, not only the giants). We also found isolated outliers at $[\teff, \logg, \feh] = [5400, 2.0, -1.5], [6900, 2.0, -1.5]$, and $[5600, 2.0, -3.0]$, which are too far from the cluster sequence to affect our analysis, and also at $[6400, 4., -2]$, $[6300, 4., -1.5]$,  $[6400, 4., -3]$. These latter outliers, and in particular the first, may affect the measurements near the TO and may be due to undiagnosed failure in the convergence of the models. Some convergence issues due to numerical instabilities where the convection layer is thin are identified in the header of the GSL files, and in some cases, the convection has been disabled and/or the $\xi$ parameter set to zero, but this affects warm giants, not the outliers detected here.

To study the discontinuity at $\teff = 5000$\,K further, we selected spectra from \RJ with $4300 < \teff < 5300$\,K and used FERRE to build $\chi^2$ maps. For each point of a \teff versus \logg grid, we linearly interpolated a spectrum in GSL at the \logg obtained in \RJ, and we computed the corresponding $\chi^2$ value. An example of the resulting $\chi^2$ map is presented in the right panel of Fig.~\ref{fig:chi2} for the spectrum ID = {\tt id000009229jd2456865p5898f000}. The map shows a step where the region with $\teff \leq 5000$\,K has a higher $\chi^2$, that is, the quality of the grid is lower in the low-temperature regime, and the temperature of the minimum $\chi^2$, $\teff = 5035$\,K has been biased toward a higher value.
The left panel presents the stacked  $\chi^2$ map for the 214 spectra with $4300 < \teff < 5300$\,K. The stacking shows that the problem is not due to an individual spectrum, and is definitely a sharp step in the $\chi^2$ space. The transition in the range $5000 < \teff < 5100$\,K is due to the linear interpolation \notep{used in the visualisation process}.

We are not certain about the origin of this discontinuity, but the GSL paper \citep{2013Husser} mentions that above 5000 K the reference wavelength used to determine the mean optical depth changes from 12\,000\,\AA{} to 5\,000\,\AA. Other discontinuities of the parameters are also mentioned at 4000\,K and 8000\,K, but because they are not in the region of interest for the present work, we did not explore them and we cannot report whether they are associated with discontinuities in the spectra. The first is a threshold above which NLTE corrections are used in the computations of some line profiles, and the second marks the temperature over which the molecular line list is limited to a few important species.

\citet{2021lancon} carried out an extensive comparison between the empirical X-Shooter Spectral Library \citep[XSL]{gonneau2020} and GSL, and found discrepancies arising below $\teff = 5000$\,K, particularly in the blue region of the spectra. It is difficult to precisely relate this degradation to the discontinuity we observe, however. There are indeed a number of reasons why the model matches become more problematic at low temperature.
We limited the range of our grids to $\teff \geq 5100$\,K in order to avoid problems with the interpolation at cooler temperatures.

\section{Interpolation artefacts}
\label{sec:interpolation_artefact}

As mentioned in Sect.~\ref{sec:anal_gsl_interpolation}, there are obvious concentrations of points at $\feh=-2.0$\,dex (a slice of the grid) in Fig.~\ref{fig:relation} where the analysis was made with GSL. The same pattern is seen in Fig.~\ref{fig:relation_ap}, with a second concentration at  $\feh=-2.25$\,dex, also a slice for the \AP grid. These patterns are particularly visible with the linear interpolation, but they also present with the quadratic interpolation (as  already noticed by \citealt{garciaperez2016}, who analysed the APOGEE data with FERRE). These are artefacts of the interpolation methods.

Whether it explicitly computes the derivatives or not, the minimisation algorithm descends the slope, and may get contradictory indications on the two sides of a slice where the derivatives are discontinuous. This may cause the solution to diverge from the slice (which is not seen in the distribution of the solutions), or to converge to the slice and produce concentration of points aligned on slices. This difficulty is \notep{inherent to any local  interpolation, regardless of its order, as they are by nature non-derivable at the nodes.}
However, with higher-order interpolation, the problem becomes smaller, certainly because the discontinuities are reduced. It is barely apparent with the quadratic interpolation and is invisible with the cubic one. A global spline interpolation would avoid this artefact.
These concentrations will often not bias the solutions statistically: although they are easily detected on graphs, they may concern only a small fraction of the solutions and affect a very limited region of the parameter space in the close vicinity of a slice of the grid.

\end{appendix}

\newpage

%
%
\end{document}